\documentclass[%
 reprint,
superscriptaddress,
 amsmath,amssymb,
 aps,pra
]{revtex4-2}

\usepackage{graphicx}
\usepackage{dcolumn}
\usepackage{bm}
\usepackage[colorlinks=true,linkcolor=blue,urlcolor=blue,citecolor=blue]{hyperref}

\usepackage{subfigure}
\usepackage{xcolor}
\usepackage{amsmath}
\usepackage{physics}
\usepackage{float}
\begin{document}

\title{Micromotors Driven by Spin-Orbit Interaction of Light: Mimicking Planetary Motion at the Microscale}

 \author{Ram Nandan Kumar}
 \email{ramnandan899@gmail.com}
 \affiliation{Department of Physical Sciences, Indian Institute of Science Education and Research Kolkata, Mohanpur-741246, West Bengal, India}

 \author{Jeeban Kumar Nayak}
 \affiliation{Department of Physical Sciences, Indian Institute of Science Education and Research Kolkata, Mohanpur-741246, West Bengal, India}

\author{Subhasish Dutta Gupta}
\email{sdghyderabad@gmail.com}
\affiliation{Department of Physical Sciences, Indian Institute of Science Education and Research Kolkata, Mohanpur-741246, West Bengal, India}
\affiliation{Tata Institute of Fundamental Research, Hyderabad, Telangana 500046, India}
\affiliation{Department of Physics, Indian Institute of Technology, Jodhpur 342030, India}

\author{Nirmalya Ghosh}
\email{nghosh@iiserkol.ac.in}
\affiliation{Department of Physical Sciences, Indian Institute of Science Education and Research Kolkata, Mohanpur-741246, West Bengal, India}

\author{Ayan Banerjee}
 \email{ayan@iiserkol.ac.in}
 \affiliation{Department of Physical Sciences, Indian Institute of Science Education and Research Kolkata, Mohanpur-741246, West Bengal, India}

\date{\today}
\date{\today}

\begin{abstract}

We introduce a new class of optical micromotors driven by the spin-orbit interaction (SOI) of light and spin-driven fluid flows leading to simultaneous rotation and revolution of the micromotors. The micromotors are essentially birefringent liquid crystal (LC) particles that can efficiently convert the angular momentum of light into high-frequency rotational motion. By tightly focusing circularly polarized Gaussian beams through a high numerical aperture (NA) objective into a refractive index (RI) stratified medium, we create a spherically aberrated intensity profile where the spinning motion of a micromotor optically trapped at the centre of the profile induces fluid flows that causes orbiting motion of the off-axially trapped surrounding particles (secondary micromotors). In addition, the interaction between the helicity of light and the anisotropic properties of the LC medium leads to the breaking of the input helicity, and drives the conversion of right to left-circular polarization and vice versa. This opposite helicity, or spin-to-spin conversion, causes the orbiting secondary micromotors to spin in certain cases (depending on their birefringence properties) as well so that the entire system of spinning primary micromotor and revolving and spinning secondary micromotors is reminiscent of planetary motion at mesoscopic scales. Our findings, supported by both theoretical modelling and experimental validation, not only advance the understanding of light-matter interactions at the microscale but also open new avenues for the design and control of next-generation micromotors and spin-orbit optomechanics, offering a versatile platform for exploring novel optical manipulation techniques in combination with complex fluid dynamics.

\end{abstract}


\maketitle

\section{Introduction}

The ability to use light to control and manipulate microscopic particles precisely has revolutionized the field of micro-robotics, unlocking unprecedented possibilities in exercising precise control over the motion of such micromotors - both translational and rotational \cite{xu2017light, palagi2019light, wani2017light, zeng2014high}. Thus - light-driven micromotors can be categorized into three primary types based on their working mechanisms \cite{knopf2018light}. The first type, opto-mechanical micromotors, harnesses the interaction between light and photosensitive materials to convert optical energy into mechanical movement \cite{zeng2018light, zeng2014high, wang2020somatosensory, lv2016photocontrol}. The second type, opto-chemical micromotors, generates propulsion forces through photochemical reactions \cite{xu2017light, aubret2018targeted, palagi2019light}. While opto-chemical micromotors show significant promise, they often depend on specific conditions, such as the availability of photoactive materials or environments with active photochemical agents \cite{villa2019fuel}. The third type, optical micromotors, achieves motion by directly manipulating microfabricated gear structures using light within optical tweezers (OT) \cite{phillips2014shape, friese2001optically, zou2020controllable}. This type is rather flexible and requires little or no sample preparation, but the requirement of gear-like microstructures to drive the secondary motors entail expensive and advanced microfabrication facilities.

The optomechanics generated by OT is driven by the intensity or polarization of light, with the rotational motion being typically induced by the latter, where angular momentum is transferred from light to the motors \cite{he1995direct,friese2001optically}. The light intensity may be used to obtain rotational motion as well, but this typically involves breaking the symmetry of the system. This can be achieved by altering the symmetry of the input beam \cite{zhao2015curved,kovalev2016optical}, modulating the beam in various manners \cite{vizsnyiczai2017light}, or introducing asymmetry in the morphology, shape, and structure of the micromotors themselves \cite{villa2019fuel}. Over the past few decades, the multifarious approaches in inducing rotation have led to substantial progress in both design as well as application of micromotors, particularly in fields ranging from biological manipulation to microfluidic systems \cite{palagi2019light, palima2013gearing, phillips2014shape, palagi2016structured, butaite2019indirect, mohanty2012optically, bunea2019strategies}. 
Among these, the spin-orbit interaction (SOI) of light, particularly in OT, has emerged as a powerful mechanism for controlling micromotors \cite{bliokh2015spin, friese2001optically, kumar2022probing}. The SOI of light, especially in non-paraxial regimes, introduces a new dimension of control by coupling the spin and orbital angular momentum of light, thereby inducing rotational motion in microscale objects without the need for physical gears or other structured elements \cite{kumar2024probing, kumar2024inhomogeneous, shao2018spin}. This interaction between light's angular momentum (AM) and the anisotropic properties of the medium can be utilized to design micromotors that demonstrate not only primary rotational motion but also complex secondary behaviors, if controlled carefully.

In this paper, we present such a new class of optical micromotors driven by the SOI of light. These micromotors are distinguished by their ability to be controlled through the helicity of light - specifically by the manipulation of the input helicity and generation of opposite helicity (or helicity breaking) of the light field. The fundamental unit of these micromotors consists of spherical birefringent mesoscopic bipolar liquid crystal (LC) micromotors \cite{prishchepa2008optical, brasselet2009optical}. These micromotors uniquely convert the angular momentum of light into mechanical motion, effectively using the SOI as a `gear' mechanism, thereby eliminating the need for traditional mechanical microfabricated structures \cite{garces2003observation}. Thus, our experimental setup involves tightly focusing a right circularly polarized (RCP) or left circularly polarized (LCP) Gaussian beam using a high numerical aperture (NA) objective lens into a stratified medium with layers having different refractive index (RI) \cite{pal2020direct}. This setup expands the beam profile beyond the Gaussian shape, crucial for trapping large birefringent particles or multiple small particles simultaneously \cite{roy2013controlled}. Our results demonstrate that the resulting helicity components of light can induce a complex array of motions, including the primary spinning of central particles in the direction dictated by the input helicity and the generation of secondary fluid flows (or spin) flows around them. One of the most intriguing outcomes of our study - where surrounding particles revolve and spin around a central spinning particle, is very much reminiscent of planetary motion, albeit in a completely different context. The spinning and revolving motions are intricately dependent on the input helicity and the birefringent properties of the particles, providing precise control over their behavior. This work establishes a novel platform for developing spin-engineered micromotors based on the SOI of light, offering new insights into light-matter interactions and paving the way for practical applications in micro-robotics, microswitches, fluid dynamics and beyond \cite{palagi2019light, villa2019fuel, butaite2019indirect}. We model the observed particle dynamics using Mueller matrix-based spin analysis, highlighting the critical role of helicity components in driving the motion of micromotors \cite{nayak2023spin}. We first lay down the theoretical premise of our work.

\begin{figure*}[!t]
    \centering
    \includegraphics[width=\textwidth]{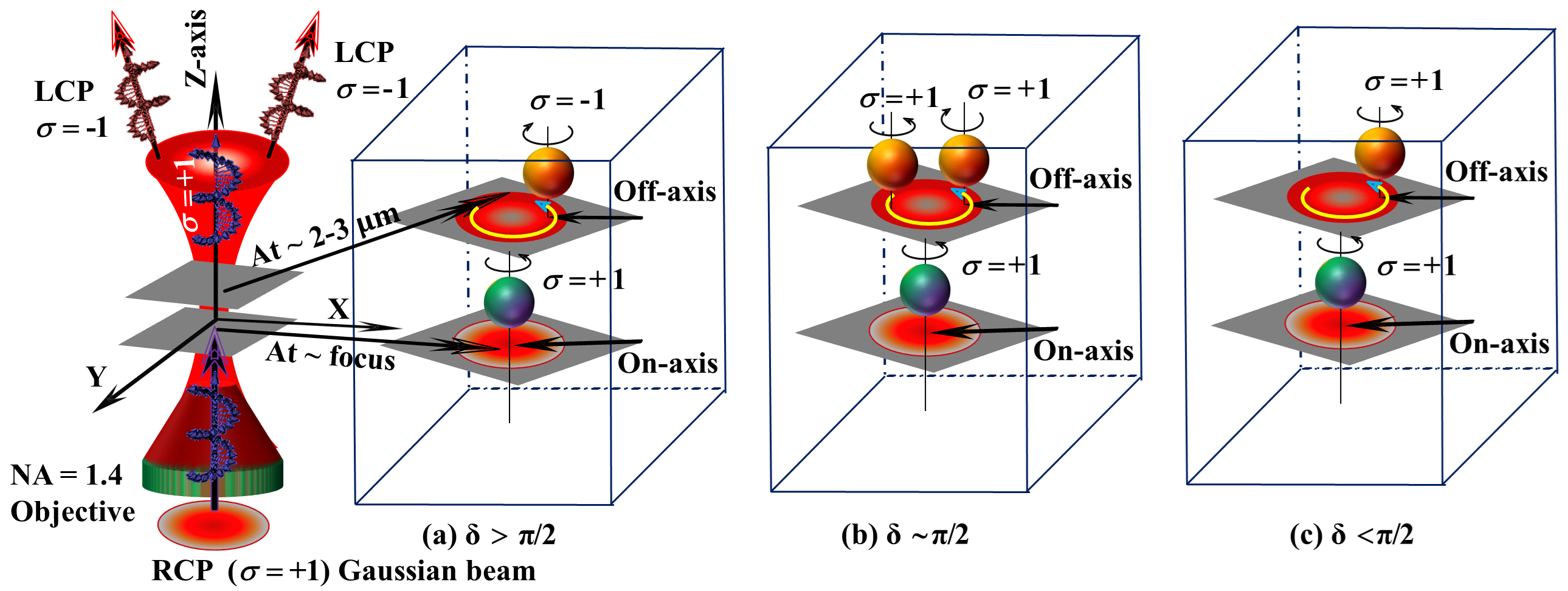}
    \caption{Schematic diagram of primary and secondary helicity ($\sigma$) generation as a consequence of the composite effect of tight focusing and anisotropic interaction of spin-polarized light. The primary helicity is always influenced by the input helicity; however, secondary helicity generation is induced by the primary helicity at off-axis positions. The sign of the helicity ($\sigma_{+}/\sigma_{-}$) depends on the value of linear retardance ($\delta$) of the centrally trapped (primary) particle. (a) For $\delta > \pi/2$, off-axis trapped particles spin in the opposite direction (induced by $\sigma_{-}=-1$) compared to the centrally trapped (primary) spinning particle ($\sigma_{+}=+1$). (b) For $\delta = \pi/2$, off-axis trapped particles may spin in the same direction ($\sigma_{+}=+1$) or the opposite direction ($\sigma_{-}=-1$), depending on their spatial location. (c) For $\delta < \pi/2$, off-axis trapped particles spin in the same direction ($\sigma_{+}=+1$), regardless of their spatial location.}
\label{schematic_machine}
\end{figure*}

\section{Theory}

Tight focusing using objective lenses with a high NA generates a non-paraxial condition. To determine the electric and magnetic fields of a circularly polarized Gaussian beam under these non-paraxial conditions, we employ the angular spectrum method or the Vector Diffraction theory of Richards and Wolf \cite{richards1959electromagnetic, Novotny2012,roy2013controlled} (detailed calculations are provided in the Appendix (\ref{App_A})). The electric field components (Ex, Ey, and Ez) of a focused circularly polarized (RCP/LCP) Gaussian beam in the focal plane, expressed in Cartesian coordinates (x, y, and z), can be described as

\begin{equation}
\left[\begin{array}{l}
E_x^0 \\
E_y^0 \\
E_z^0
\end{array}\right]_{R C P / L C P}^{Gauss}=\left[\begin{array}{l}
I_{00}+I_{02} \cos 2 \psi \pm i I_{02} \sin 2 \psi \\
I_{02} \sin 2 \psi \pm i\left(I_{00}-I_{02} \cos 2 \psi\right) \\
-2 i I_{01} \cos \psi \pm 2 I_{01} \sin \psi
\end{array}\right]
\label{eq1}
\end{equation}

Here, $\mathbf{E}^0$ denotes the output electric fields, with the suffix ${Gauss}$ indicating the Gaussian beam. The terms $I_{00}$, $I_{01}$, and $I_{02}$ represent the Debye-Wolf integrals (the detailed expression is provided in the Appendix (\ref{App_A})) \cite{roy2013controlled, kumar2022probing}, and $\psi$ is the azimuthal angle in the cylindrical (or spherical) coordinate system. It is important to note that a circularly polarized Gaussian beam does not carry orbital angular momentum (OAM) but does possess spin angular momentum (SAM) with a magnitude of $\pm\hbar$ per photon, so the total angular momentum equals the SAM before focusing. However, after the tight focusing of the RCP/LCP Gaussian beam, opposite (or orthogonal) helicity and longitudinal components emerge, along with the corresponding OAM modes in the output electric field, as a consequence of  SOI \cite{bliokh2015spin}. We now decompose the output electric fields given in Eq.~(\ref{eq1}) for an input RCP or LCP Gaussian beam into their respective SAM and OAM components as

\onecolumngrid
 
\begin{equation}
\left[\begin{array}{c}
E_x^0 \\
E_y^0 \\
E_z^0
\end{array}\right]_{R C P / L C P}^{G a u s s}=I_{00}\left[\begin{array}{c}
1 \\
\pm i \\
0
\end{array}\right] \pm I_{02} \exp ( \pm 2 i \psi)\left[\begin{array}{c}
1 \\
\mp i \\
0
\end{array}\right]-2 i I_{01} \exp ( \pm i\psi)\left[\begin{array}{l}
0 \\
0 \\
1
\end{array}\right]
\label{eq2}
\end{equation}

\twocolumngrid
From Eq.~(\ref{eq2}), it is clear that the coefficient of the first term, \(I_{00}\), corresponds to the helicity of the input RCP/LCP polarized Gaussian beam (same helicity).
The transfer of this helicity component to the birefringent particle induces spinning in the direction of the circulating electric field (see Fig.~\ref{schematic_machine} (a)-(c)) - which is observed for the micromotor trapped at the beam center. The coefficient of the second term, \(I_{02}\), associated with an OAM of \(\pm 2\hbar\), represents the flipping of helicity (\(\sigma = \pm1 \rightarrow \mp1\), i.e., opposite helicity). The coefficient of the third term, \(I_{01}\), associated with an OAM of \(\pm \hbar\), describes the conversion from circular to linear polarization (\(\sigma = \pm1 \rightarrow 0\)), and represents the longitudinal component of the electric field. The coefficients of the last two terms in Eq.~(\ref{eq2}) determine the strength of the SOI in the non-paraxial regime (plots of the squared moduli of these coefficients are provided in the Appendix (\ref{APP_B})). It is important to note that spin-to-orbit conversion occurs due to the adiabatic evolution of the \( k \)-vector (momentum vector) and the polarization of light during the focusing process, which leads to the spin-redirection geometric Berry phase \cite{gupta2015wave,roy2013controlled}. The gradient of this geometric Berry phase results in the emergence of OAM (or vorticity).

On a different note, the interaction between a spin-polarized (RCP/LCP) focused Gaussian beam and an anisotropic medium (birefringent LC particles) leads to the generation of light of opposite helicity. Consequently, the interaction of the tightly focused light with the micromotor trapped at the beam center, causes the generation of light of opposite helicity for certain physical properties of the birefringent micromotor. The out-coupled light now determines the motion of the micromotor(s) trapped off-axis and at higher axial distance (from the trap focus) from the central micromotor. Thus, the centrally trapped micromotors (LC particles/droplets) work as q-plates (or linear retarders), with spatially azimuthally varying orientation of the directors \cite{marrucci2006optical,kumar2024rectangular}. Then, the final expression for the output electric field emerging from the central trap micromotor can be written as

\begin{equation}
   \begin{aligned}
\left(\begin{array}{c}
1 \\
\sigma i
\end{array}\right) \longmapsto \cos \delta / 2 \cdot\left(\begin{array}{c}
1 \\
\sigma i
\end{array}\right)-i e^{2 \sigma i \psi} \cdot \sin \delta / 2\left(\begin{array}{c}
1 \\
-i \sigma
\end{array}\right) 
\label{eq3}
\end{aligned} 
\end{equation}


Here, $\sigma$ denotes the helicity of the field, with values ranging between -1 and 1 ($-1 \leq \sigma \leq 1$). For completely circularly polarized LCP and RCP light, $\sigma$ takes the values of -1 and +1, respectively. $\psi$ represents the orientation angle of the director of the micromotor (LC particle), and $\delta$ denotes the linear retardance of the central trap LC particle. The first term of Eq.~(\ref{eq3}) represents the component of the same helicity, with its magnitude determined by \( \cos(\delta/2) \). However, the second term of Eq.~(\ref{eq3}) illustrates the component of the opposite (or orthogonal) helicity, with its magnitude determined by \( \sin(\delta/2) \). Now, it is important to note the following consequences of this term: 

\begin{enumerate}
    \item If \(\delta > \pi/2\), the off-axis trapped LC particle will follow the opposite (or orthogonal) helicity with an OAM mode, and its magnitude is given by \(\sin(\delta/2)\), as shown in schematic Fig.~\ref{schematic_machine}(a).
    
    \item If \(\delta = \pi/2\), the first and second terms of Eq.~(\ref{eq3}) contribute equally. As a result, the spin response of the secondary LC particle will depend on its off-axis spatial location, allowing it to follow the same or opposite helicity of the field, as illustrated in the schematic Fig.~\ref{schematic_machine}(b).
    
    \item If \(\delta < \pi/2\), the off-axis trapped LC particle will follow the input (or the same) helicity of the field and its magnitude is given by \(\cos(\delta/2)\), as shown in schematic Fig.~\ref{schematic_machine}(c).
\end{enumerate} 

We now proceed to carry out a numerical simulation of our experimental system in the light of the theory presented above, and develop intuitions about possible experimental observations. 
\begin{center}
\begin{figure}[!h]
\includegraphics[width=0.5\textwidth]{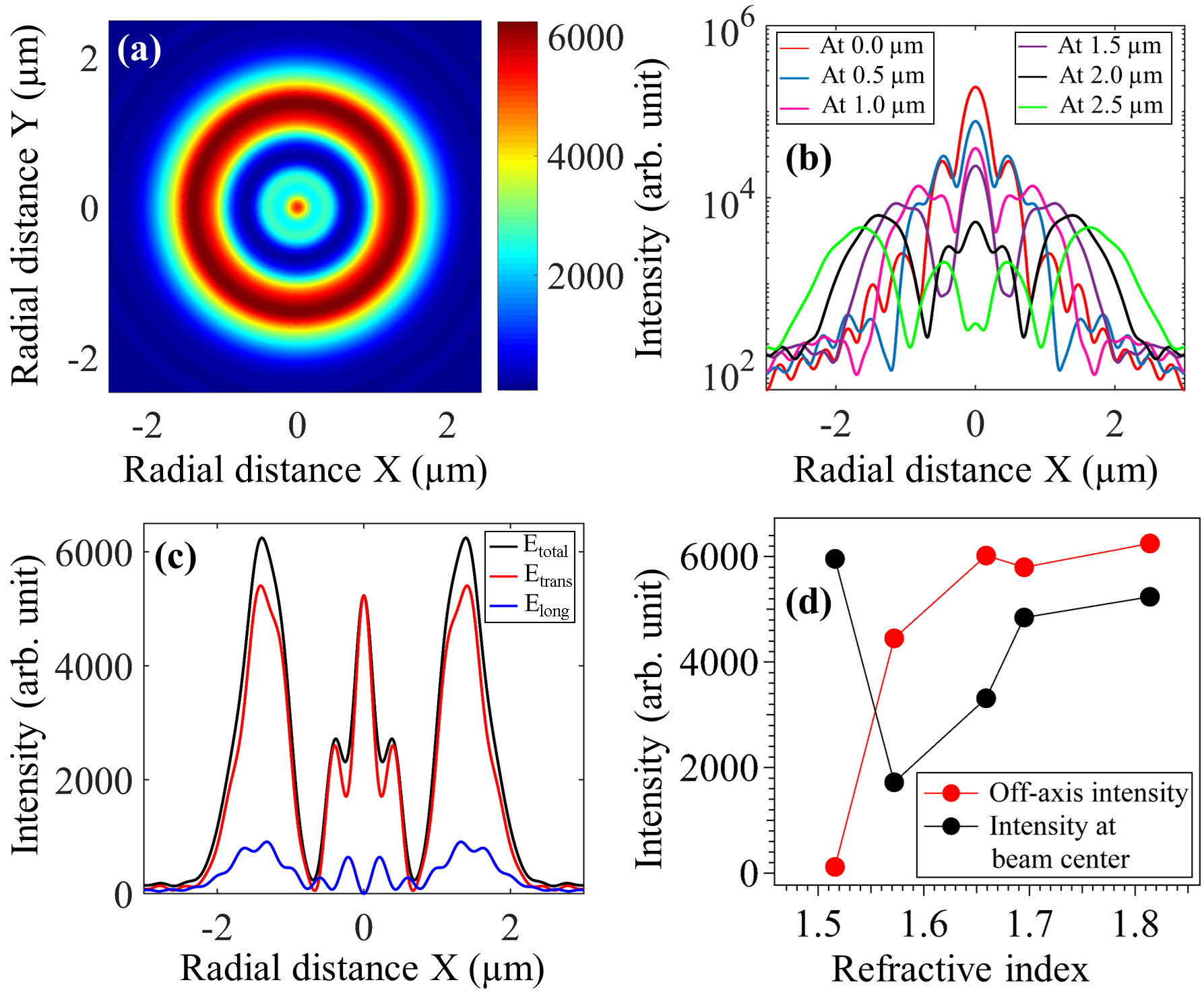}
\caption{Numerical analysis of the intensity distribution for a mismatch RI 1.814 at $z = 2 \mu\text{m}$ from the focus of a high NA objective lens for an input circularly polarized Gaussian beam. (a) Intensity distribution at $z = 2 , \mu\text{m}$ from the focus. (b) Comparison of intensities at the trap center (solid black circles) and off-axis (solid red circles) as a function of the RI at $2 \mu\text{m}$ from the focus. (c) Comparison of the transverse, longitudinal, and total electric field intensity distributions (xy) of a circularly polarized Gaussian beam for a mismatched RI (1.814) at $2 \mu\text{m}$ from the focus. (d) Line plot of the intensity distribution (in log scale) of a circularly polarized Gaussian beam at different $z$ planes for a mismatched RI (1.814), showing how the Gaussian nature of the beam profile diminishes as we move away from the focus.}
\label{intensity} 
\end{figure}
\end{center}

\section{Numerical simulations}\label{Num_simulations}

In our experimental system, the output beam from the high NA objective lenses in the optical tweezers setup is passed through a stratified medium. The laser beam of wavelength 671 nm is incident on the 100X oil immersion objective of NA 1.4 followed by (a) an oil layer of thickness around 5 $\mu m$ and RI 1.516, (b) a 160 $\mu m$ thick coverslip having RI varying between 1.516-1.814 (note that the case where the $RI = 1.516$ is henceforth referred to as the ``matched condition,” which is typically employed in optical tweezers to minimize spherical aberration effects in the focused beam spot, whereas the other values are referred to as a `mismatched' condition) (c) a sample chamber of an aqueous solution of LC particles/droplets in a water medium having a RI of 1.33 with a depth of 35 $\mu m$, and finally (d) a glass slide of RI 1.516 whose thickness we consider to be semi-infinite ( 1500 $\mu m$) [see Fig.~\ref{schematic} (a)]. In the simulation, the origin of coordinates is taken inside the sample chamber at an axial distance of 5 $\mu m$ from the interface between the sample and the coverslip. Thus, the objective-oil interface is at -170 $\mu m$, the oil-cover slip interface is at -165 $\mu m$, the cover slip-sample chamber interface is at -5 $\mu m$, and the sample chamber-glass slide interface is at +30 $\mu m$. At the paraxial to non-paraxial transformation boundary, we decompose the incident polarized rays $E_{inc}$ into s-polarization and p-polarization. For each polarization state, we account for the corresponding Fresnel transmission coefficients $T_{s}$ and $T_{p}$, as well as the reflection coefficients $R_{s}$ and $R_{p}$ at each interface of the RI-stratified medium \cite{Novotny2012,roy2013controlled,kumar2022probing,kumar2024probing}. 

\subsection{Numerical simulations of intensity distribution}

Fig. ~\ref{schematic} (a) is a cartoon representation of our numerical and experimental system. Following the proposed strategy, we perform numerical simulations to demonstrate the radial variation of intensity distribution at 2 $ \mu m$ above the focal plane for an RI 1.814. In Fig.~\ref{intensity} (a)-(d), we show the intensity distribution of a circularly polarized (RCP/LCP) Gaussian beam. For the matched condition of RI 1.516, the intensity distribution is purely Gaussian. However, due to the mismatched RI (1.814) of the coverslip, the spherical abbrated intensity distribution is no longer Gaussian (see Fig.~\ref{intensity} (a)). In Fig.~\ref{intensity}(b), the line plot of the intensity distribution of circularly polarized light at different z-planes (axial-planes) for mismatched RI (1.814) shows that as we move away from the focus, off-axis intensity lobes are formed. Hence, we get enough spatial extent in the transverse plane to trap more than one particle at a time. In Fig.~\ref{intensity}(c), we separately plot the transverse and longitudinal components of the electric field intensity distribution and compare them with the total electric field intensity. We observe that the intensity corresponding to the longitudinal component of the electric field (\( E_\text{long} \)) is primarily concentrated at the off-axis position due to the first-order Bessel function (\( J_{1} \)) embedded in the \( I_{01} \) coefficient, which accounts for around 15\% of the total intensity. However, the distribution of the transverse intensity profile $E_{\text{trans}}$ (arising from $I_{00}$ and $I_{02}$) is concentrated at both the beam center and off-axis, leading to an annular intensity ring (the zero-order Bessel function  $J_{0}$ embedded in $I_{00}$ has a non-vanishing value at the on-axis). The central maxima, however, has around 84\% of the total intensity as illustrated in Fig.~\ref{intensity}(c). Therefore, when an LC particle is trapped at the beam center, it will follow the input helicity of the light as the $I_{00}$ coefficient in Eq.~(\ref{eq2}) is associated with the input RCP/LCP helicity). In order to obtain the most optimized experimental conditions, we plotted the variation of intensity at the beam center (on-axis) and off-axis positions as a function of the RI in Fig. ~\ref{intensity} (d). It is noteworthy that under the matched condition of RI 1.516, at 2 $\mu m$ away from the focus, most of the intensity is concentrated at the beam center, with less intensity distributed off-axis. This indicates that the Gaussian beam retains its Gaussian nature. However, under the mismatched condition of RI 1.814, the intensity at the beam center (on-axis) and off-axis positions (annular intensity ring) become comparable. The resulting intensity profile resembles that of a tightly focused first-order radially polarized vector beam, as demonstrated in refs. ~\cite{kumar2022manipulation,kumar2024probing,kumar2024inhomogeneous}. This configuration enables particles to be trapped in both the central and annular regions.

\begin{center}
\begin{figure}[!t]
\includegraphics[width=0.5\textwidth]{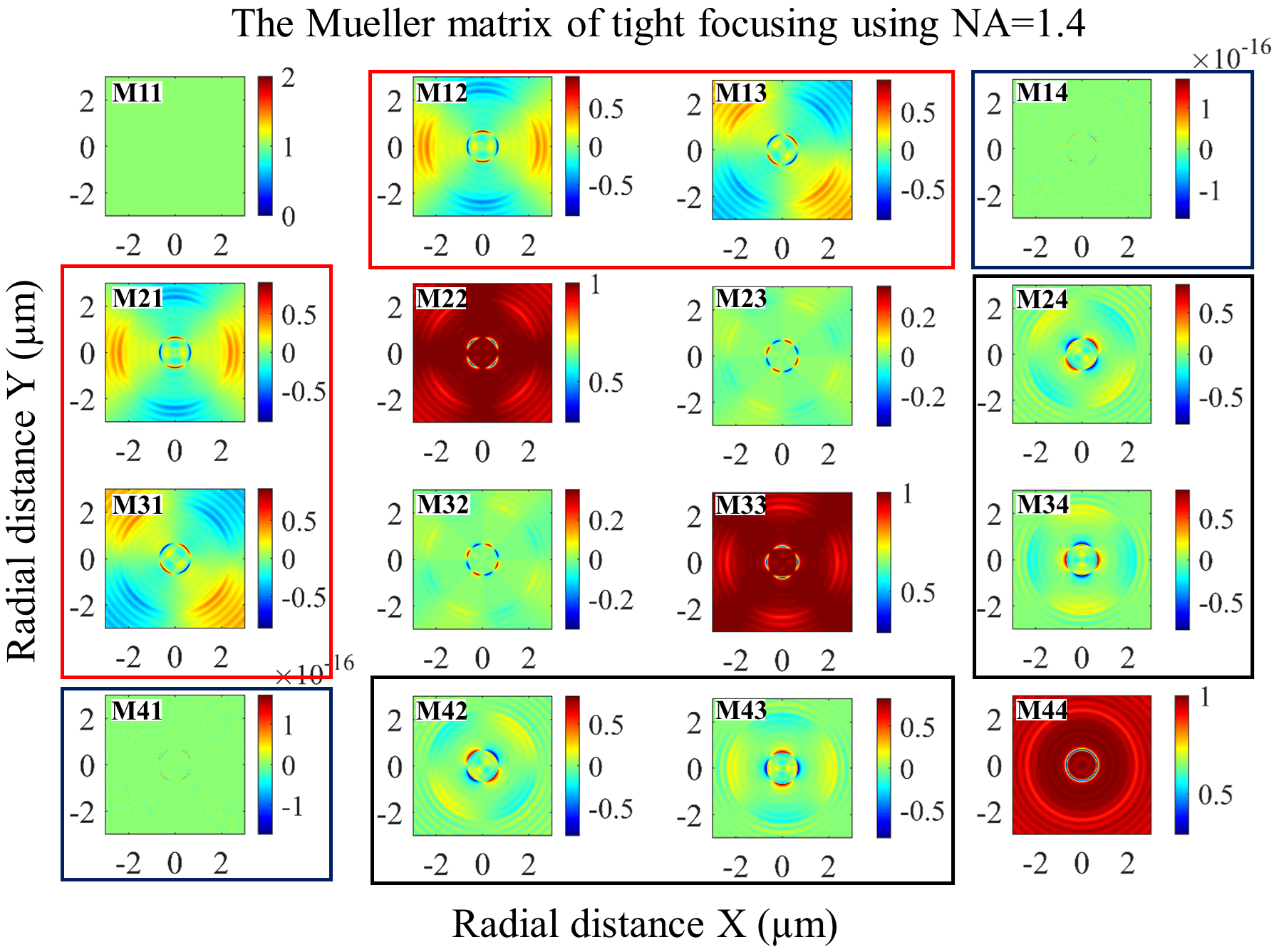}
\caption{Numerically computed 4x4 Mueller matrix for the tight focusing of a Gaussian beam. The signature of the linear diattenuator is indicated by elements M\(_{12}\) = M\(_{21}\) \(\propto \cos(2\psi)\) and M\(_{13}\) = M\(_{31}\) \(\propto \sin(2\psi)\) (marked by red boxes). The linear retarder is indicated by elements M\(_{24}\) = -M\(_{42}\) \(\propto -\sin(\delta) \sin(2\psi)\) and M\(_{34}\) = -M\(_{43}\) \(\propto \sin(\delta) \cos(2\psi)\) (marked by black boxes). The elements M\(_{14}\) and M\(_{41}\) = 0, which represent circular diattenuation (or circular anisotropy), are highlighted with blue boxes. The elements M\(_{23}\) and M\(_{32}\) illustrate the gradient of the geometric phase and are \(\propto \cos(2\psi) \sin(2\psi) (1 - \cos\delta)\). The diagonal element M\(_{11}\) represents the total intensity of the focused field, while the remaining elements M\(_{22}\), M\(_{33}\), and M\(_{44}\) describe the depolarization parameters.}
\label{tight focusing}
\end{figure}
\end{center}

\subsection{Mueller matrix of tight focusing}

To analyze the SOI effect using conventional polarization parameters such as diattenuation (\(d\)) and retardance (\(\delta\)), we derive the Mueller matrix corresponding to the Jones matrix of tight focusing. Diattenuation represents the differential attenuation of orthogonal polarizations, while retardance accounts for the phase anisotropy, i.e., the phase difference introduced between orthogonal polarization states. These polarization effects relate to the amplitude and phase components of the focused field, respectively. The Mueller matrix corresponding to the 2x2 Jones matrix \(J\) (extracted from the first two rows and columns of the 3x3 Jones matrix in Eq.~(\ref{Eqa_4})) can be derived using the standard formula \(M = \mathbf{\mathcal{A}} \cdot \left(J \otimes J^*\right) \cdot \mathbf{\mathcal{A}}^{-1}\). Here, \(J\) is the 2x2 Jones matrix for tight focusing, and \(\mathbf{\mathcal{A}}\) is a 4x4 matrix defined as  \cite{lu1995interpretation,gupta2015wave}
\begin{equation}
J=\left[\begin{array}{cc}
I_{00}+I_{02} \cos 2 \psi & I_{02} \sin 2 \psi \\
I_{02} \sin 2 \psi & I_{00}-I_{02} \cos 2 \psi
\end{array}\right],
\label{Eqa_10}
\end{equation}

\begin{equation*} 
\mathbf{\mathcal{A}}=\left(\begin{array}{cccc}
1 & 0 & 0 & 1 \\
1 & 0 & 0 & -1 \\
0 & 1 & 1 & 0 \\
0 & -i & i & 0
\end{array}\right),
\end{equation*}

The resulting matrix is a diattenuating retarder Mueller matrix, \( M_{TF}(d_{TF}, \delta_{TF}, \psi) \), characterized by diattenuation \( d_{TF} \), retardance \( \delta_{TF} \), and azimuthal angle \( \psi \), which represents the orientation of the axes of the diattenuating retarder. The subscript TF denotes the tight focusing. In Fig.~(\ref{tight focusing}), we present the numerically computed 4x4 Mueller matrix for an azimuthal diattenuating retarder. The matrix elements that represent the linear diattenuator are M\(_{12}\) and M\(_{21}\), which are proportional to \(\cos(2\psi)\), and M\(_{13}\) and M\(_{31}\), which are proportional to \(\sin(2\psi)\); these are marked with red boxes. The linear retarder is characterized by elements M\(_{24}\) and M\(_{42}\), which are proportional to \(-\sin(\delta) \sin(2\psi)\), and M\(_{34}\) and M\(_{43}\), which are proportional to \(\sin(\delta) \cos(2\psi)\); these elements are highlighted with black boxes. Elements M\(_{14}\) and M\(_{41}\), which are highlighted with blue boxes, are equal to zero and represent circular diattenuation (or circular anisotropy). Additionally, elements M\(_{23}\) and M\(_{32}\) illustrate the gradient of the geometric phase, being proportional to \(\cos(2\psi) \sin(2\psi) (1 - \cos\delta)\). The diagonal element M\(_{11}\) represents the total intensity of the focused field, while the other diagonal elements M\(_{22}\), M\(_{33}\), and M\(_{44}\) denote the depolarization parameters.

\begin{figure}[!t]
    \centering
    \includegraphics[width=0.5\textwidth]{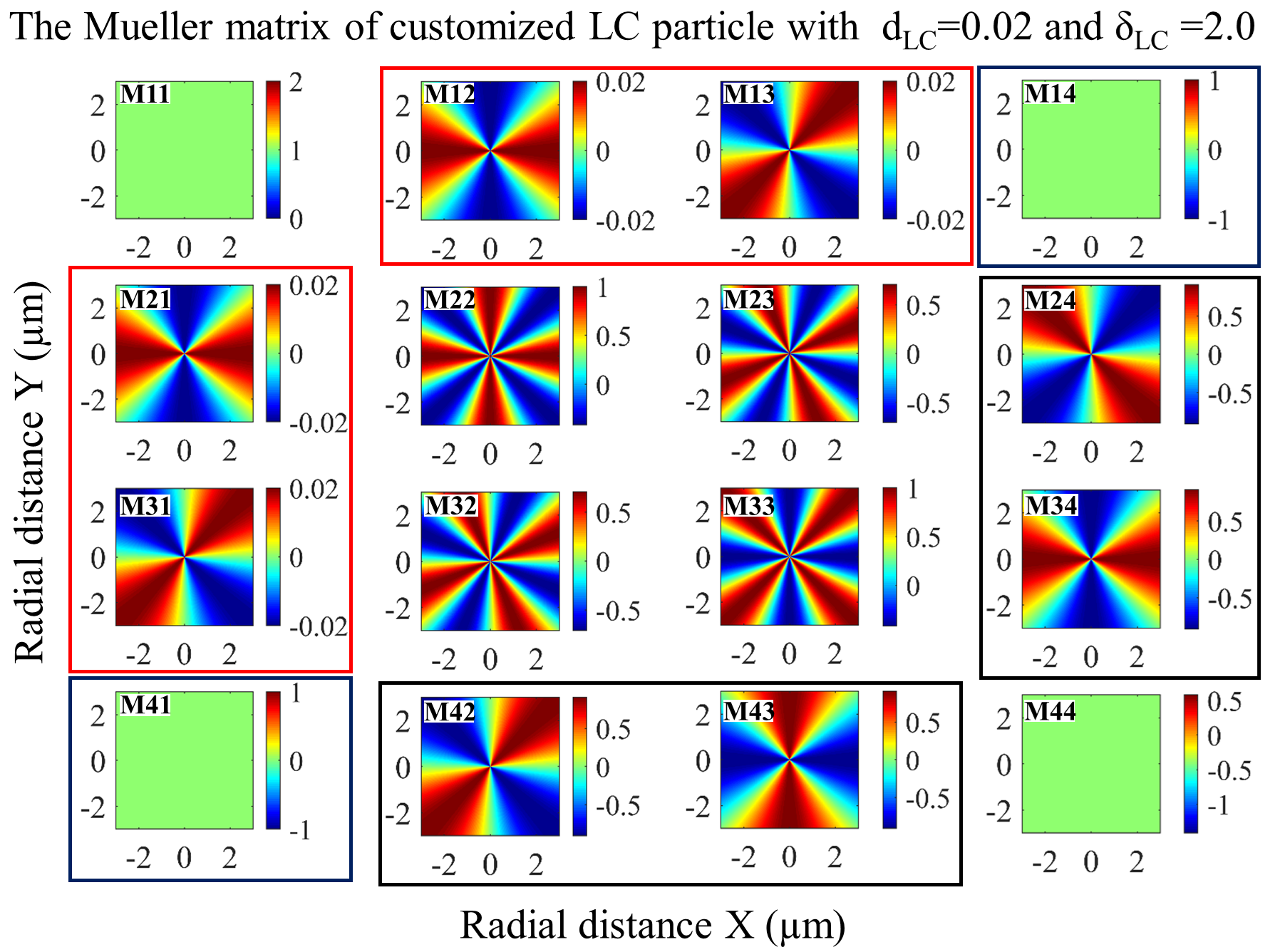}
    \caption{The Mueller matrix of a customized LC particle. The signature of the linear diattenuator is indicated by elements M\(_{12}\), M\(_{13}\), M\(_{21}\), and M\(_{31}\), marked by red boxes. The linear retarder is represented by elements M\(_{24}\), M\(_{34}\), M\(_{42}\), and M\(_{43}\), marked by black boxes. Elements M\(_{14}\) and M\(_{41}\), highlighted in blue boxes, describe circular diattenuation; however, their values are zero for LC particles. The M\(_{44}\) element shows a constant value.}
\label{MM of customized LC}
\end{figure}

\subsection{Mueller matrix of customized LC particles}

The general Mueller matrix elements (M$_{ij}$, where \( i \) represents the row and \( j \) the column, with \( i, j = 1, 2, 3, 4 \)) corresponding to a diattenuating retarder are given as\cite{hilfiker2004mueller,zaidi2024metasurface,soni2013enhancing,lu1996interpretation,gupta2015wave}

\begin{equation}
\begin{aligned}
& M_{11}=1;\quad M_{12}=M_{21}=d \cos 2 \psi;\quad M_{13}=M_{31}=d \sin 2 \psi ; \\
& M_{14}=M_{41}=0 ; \quad M_{22}=\cos^2 2 \psi + x \cos \delta \sin^2 2 \psi ; \\
& M_{23}=M_{32}=\sin 2 \psi \cos 2 \psi - x \cos \delta \sin 2 \psi \cos 2 \psi ; \\
& M_{24}=-M_{42}=-x \sin \delta \sin 2 \psi ; \\
& M_{33}=\sin^2 2 \psi + x \cos \delta \cos^2 2 \psi ; \\
& M_{34}=-M_{43}=x \sin \delta \cos 2 \psi ; \\
& M_{44}=x \cos \delta ; \quad x=\left|\sqrt{1-d^2}\right| .
\end{aligned}
\label{Eqa_9}
\end{equation}

Here, \( d \) represents the diattenuation, \( \delta \) is the retardance, and the azimuthal angle \( \psi \) corresponds to the orientation of the axes of the diattenuating retarder. We first quantified the polarization parameters, mainly diattenuation (\( d \)) and retardance (\( \delta \)), of the LC particle using the Lu-Chipman (polar) decomposition method  \cite{xing1992deterministic, lu1995interpretation, nayak2023spin}, with detailed information provided in the experimental section (\ref{exp_methods}) (see Figs.~\ref{ret} (a)-(d)). The experimentally observed values of \( d = 0.02 \) and \( \delta = 2.0 \) were then substituted into Eq.~(\ref{Eqa_9}), allowing us to numerically compute the Mueller matrix of the customized LC particle, \( M_{LC}(d_{LC},\delta_{LC},\psi) \), as shown in Fig.~(\ref{MM of customized LC}). Experimentally, we collected multiple sets of data and observed that our sample exhibited a relatively constant diattenuation value of around \( d = 0.02 \). However, the linear retardance (\(\delta_{LC}\)) values varied between approximately 1 and 3. We categorized these into three distinct groups: \(\delta_{LC} < \pi/2\), \(\delta_{LC} = \pi/2\), and \(\delta_{LC} > \pi/2\). Here, we present the results corresponding to \(\delta_{LC} > \pi/2\) in Fig.~(\ref{MM of customized LC}), noting that for the other two cases (\(\delta_{LC} < \pi/2\) and \(\delta_{LC} = \pi/2\)), the distribution of the Mueller matrix elements (M$_{ij}$) remains similar, with differences only in the maximum and minimum values.

\section{Results and discussion} \label{res_dis}

\subsection{Resultant Mueller matrix}

The propagation of a spin-polarized Gaussian beam from a cylindrical to a spherical symmetric system through an aplanatic lens (i.e., the paraxial to non-paraxial transformation) generates SOI of light. This SOI effect, or spin-to-orbital angular momentum coupling, is described by Eq.~(\ref{eq2}). The spin-to-orbital coupling (or SOI) is intrinsic in nature and is mediated by the azimuthally varying geometric phase \( e^{i\ell \psi} \), which evolves due to the azimuthal trajectory of the polarized field (spin redirection Berry phase). Additionally, when tightly focused spin-polarized light interacts with the anisotropic medium of an LC particle, a geometric phase (or Pancharatnam-Berry phase) is generated due to the bipolar variation of the anisotropy axis of the LC particle (or linear retarder) \cite{bliokh2015spin, gupta2015wave, nayak2023spin}. Here, the tight focusing of the beam and the interaction of focused spin-polarized light with the anisotropic medium of the LC particles together create a composite effect. To investigate this phenomenon further, we numerically computed the 4x4 resultant Mueller matrix (MM) for the tight focusing of a Gaussian beam on a bipolar LC particle \cite{kumar2024inhomogeneous}. Initially, we numerically computed the 4x4 MM ($M_{TF}(d_{TF},\delta_{TF},\psi)$) for the tight focusing of the Gaussian beam using a 2x2 Jones matrix of the transverse electric field, which revealed a characteristic form of a linear diattenuator $d_{TF}$ (partially polarizing), and a retarder $\delta_{TF}$ (wave plate) with azimuthal orientation $\psi$. Following this, we experimentally recorded the 4x4 MM data for the bipolar LC particle (detailed procedures are mentioned in the Experiments and Methods section (\ref{exp_methods})).

\begin{figure}[!t]
\centering
\includegraphics[width=0.5\textwidth]{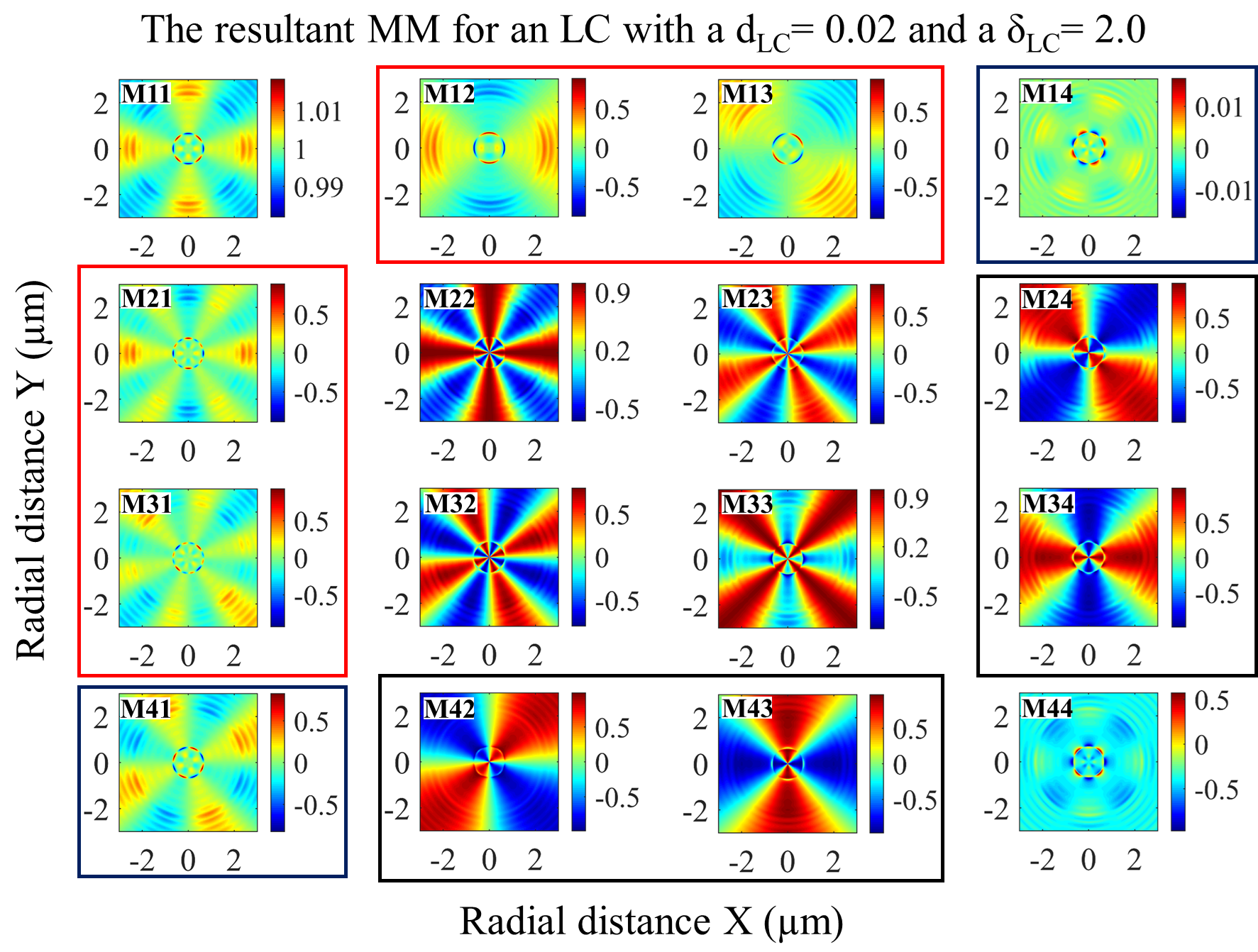}
\caption{Numerically computed 4x4 resultant Mueller matrix for the tight focusing of a Gaussian beam on an LC particle with linear diattenuation $d_{LC}=0.02$ and retardance $\delta_{LC}=2$. The signature of the linear diattenuator is indicated by elements M${12}$, M${13}$, M${21}$, and M${31}$ (marked by red boxes). The linear retarder is indicated by elements M${24}$, M${34}$, M${42}$, and M${43}$ (marked by black boxes). The elements M${14}$ and M${41}$, which describe the Spin Hall Effect (SHE), i.e., the difference in RCP and LCP components of the electric field in the focal plane, are highlighted in blue boxes.}
\label{MM}
\end{figure}

Using the LU-Chipman (or polar) decomposition \cite{lu1995interpretation,nayak2023spin}, we quantified the polarization parameters, including linear diattenuation ($d_{LC}$) and linear retardance ($\delta_{LC}$), of the LC particles (described in detail in Section (\ref{exp_methods}) see Fig.~\ref{ret} (a)-(d)). We collected multiple data sets from different particles and observed that our sample exhibited a range of linear retardance values ($\delta_{LC}$) from around 1 to 3, which we categorized into three distinct groups: $\delta_{LC} < \pi/2$, $\delta_{LC} = \pi/2$, and $\delta_{LC} > \pi/2$. From there, we numerically computed the MM of customized bipolar LC particles $M_{LC}(d_{LC},\delta_{LC},\psi)$ (see Fig.~(\ref{MM of customized LC})). The parameter $\psi$, representing the bipolar orientation of the anisotropy axis of LC molecules, generates a phase difference in the orthogonal components of the input polarization. Specifically, linear retardance and linear diattenuation arise from differences in the real and imaginary parts of the refractive index between two orthogonal linear polarization states, respectively\cite{gupta2015wave}. Hence, the resultant MM $M_{res}(d,\delta,\psi)$ of the composite effects can be modeled by the sequential product of the MM of tight focusing $M_{TF}$ and the MM of bipolar LC $M_{LC}$, representing two successive polarization-transforming events, i.e., $M_{res}(d,\delta,\psi)=M_{TF}(d_{TF},\delta_{TF},\psi) \cdot M_{LC}(d_{LC},\delta_{LC},\psi)$. Here, we show the resultant MM at $\delta>\pi/2$, while the rest of the results corresponding to $\delta_{LC}<\pi/2$ and $\delta_{LC}=\pi/2$ are available in the Appendix (\ref{APP_B}).

In Fig.~(\ref{MM}), we plotted the resultant Mueller matrix (MM) at \(\delta > \pi/2\) for the composite effect, highlighting the linear retarder elements (M\(_{24}\), M\(_{34}\), M\(_{42}\), and M\(_{43}\)) with black boxes, which indicate the signature of SAM to intrinsic OAM (vortex) conversion due to an azimuthal linear retarder. The linear diattenuator elements, marked by red boxes (M\(_{12}\), M\(_{13}\), M\(_{21}\), and M\(_{31}\)), describe the SOI effects caused by a vortex linear diattenuator. The intensity lobes in the linear retarder and linear diattenuator elements are proportional to \(\cos{2\psi}\) or \(\sin{2\psi}\) (see Eq.~(\ref{Eqa_9})). The elements M\(_{23}\) and M\(_{32}\) illustrate the evolution of the azimuthal geometric phase (i.e., vortex phase). Additionally, the M\(_{41}\) and M\(_{14}\) elements, indicated by blue square boxes, partially represent circular anisotropy (or SOI effects). The positive and negative values of the M\(_{41}\) and M\(_{14}\) elements correspond to the positive (\(\sigma_{+}\)) and negative (\(\sigma_{-}\)) helicity of the output fields. We then proceed to calculate the resultant Stokes vector using the relation \(\mathbf{S}^0 = M_{\text{res}} \mathbf{S}^{\mathbf{in}}\), where \(M_{\text{res}}\) is the 4x4 Mueller matrix representing the composite effect, and \(\mathbf{S}^{\mathbf{in}}\) and \(\mathbf{S}^0\) are the input and output Stokes vectors, respectively.

\begin{figure*}[!t]
\centering
\includegraphics[width=\textwidth]{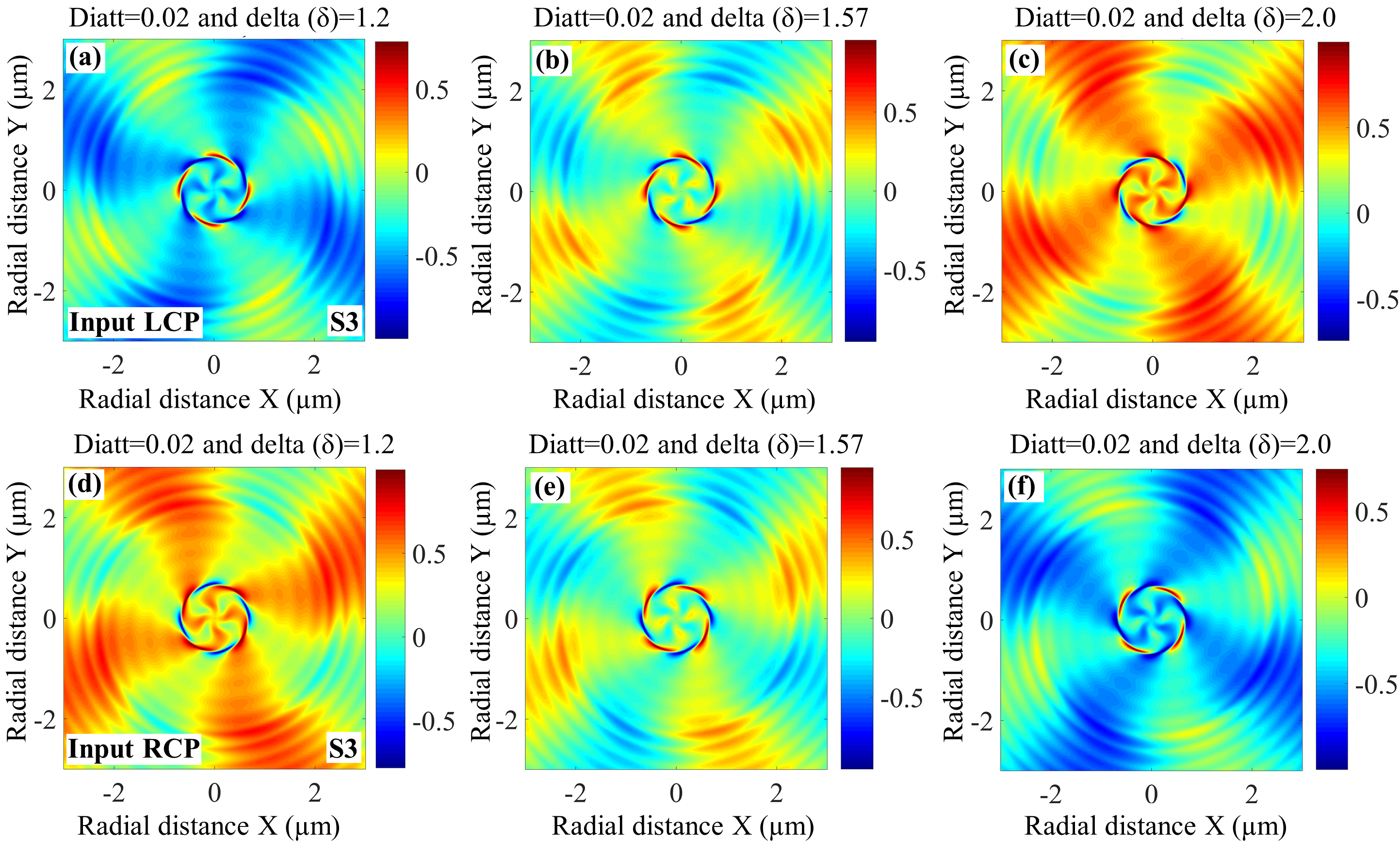}
\caption{The resultant S${3}$ component of Stokes vector elements for tightly focused circularly polarized light on an LC particle at a fixed linear diattenuation value of 0.02 and different linear retardance values. (a), (b), and (c) show the results for left circularly polarized light at 1.2, 1.57, and 2.0, respectively. (d), (e), and (f) show the results for right circularly polarized light at 1.2, 1.57, and 2.0, respectively.}
\label{Stokes vector}
\end{figure*}

\subsection{Resultant Stokes vector} 

The study of SAM and its relationship with the polarization of light reveals fascinating dynamics in tightly focused beams on LC particles. The SAM ($\mathbf{S}$), which is intrinsic in nature, depends entirely on the polarization of the beam and can be described by the equation: $\mathbf{S} \propto \operatorname{Im}\left[\epsilon\left(\mathbf{E}^* \times \mathbf{E}\right)+\mu\left(\mathbf{H}^* \times \mathbf{H}\right)\right]$, with $\epsilon$ as the electric permittivity and $\mu$ as the magnetic permeability  \cite{garces2003observation,aiello2015transverse,kumar2024probing}. The S${3}$ component of the Stokes vector, which illustrates the difference between right and left circular polarization (or the difference in $\sigma_{+}$ and $\sigma_{-}$ helicity of light), is proportional to $i\left(\left\langle E_y E_x^*\right\rangle-\left\langle E_x E_y^*\right\rangle\right)$, where $\left\langle...\right\rangle$ represents the temporal average of the electric field \cite{gupta2015wave}. Therefore, the longitudinal SAM (LSAM) density and the S${3}$ component of the Stokes vector are equivalent.

Following the experimental results, we calculated the resultant Stokes vector elements S${0}$, S${1}$, S${2}$, and S${3}$ for input LCP ($S_{LCP}^{in}=\left[\begin{array}{llll} 1 & 0 & 0 & -1 \end{array}\right]^{T}$) and RCP ($S_{RCP}^{in}=\left[\begin{array}{llll} 1 & 0 & 0 & 1 \end{array}\right]^{T}$) light using the resultant MM of the composite effect $M_{\text{res}}$ (tight focusing of a Gaussian beam on an LC particle) at a fixed linear diattenuation value of 0.02 and different linear retardance ($\delta_{LC}$) values (see Figs.~(\ref{MM}) and \ref{MM at 1.2 and 1.57}). To analyze the effect of LSAM (same and opposite helicity of light), we plotted the resultant S${3}$ component of the Stokes vector for input LCP and RCP light at linear retardance values $\delta_{LC}<\pi/2$, $\delta_{LC}=\pi/2$, and $\delta_{LC}>\pi/2$. Here, we present only the resultant S${3}$ component of the Stokes vector, while the other resultant Stokes vector elements corresponding to $\delta_{LC}<\pi/2$, $\delta_{LC}=\pi/2$, and $\delta_{LC}>\pi/2$ are detailed in Appendix (\ref{APP_B}).

In Figs.~\ref{Stokes vector} (a), (b), and (c), we show the resultant S${3}$ component of the Stokes vector for tightly focused LCP light on an LC particle at a fixed linear diattenuation  $d_{LC}$ value of 0.02 and linear retardance $\delta_{LC}$ values of 1.2, 1.57, and 2.0, respectively. Similarly, Figs.~\ref{Stokes vector} (d), (e), and (f) present the resultant S${3}$ component of the Stokes vector for tightly focused RCP light on an LC particle under the same conditions. In our convention, positive ($\sigma_{+}$) and negative ($\sigma_{-}$) values of S${3}$ represent clockwise and counter-clockwise spinning motion of the LC particle about the z-axis, respectively.

We thus find the following, in corroboration with our theoretical predictions:
\begin{enumerate}
    \item When the linear retardance value of the centrally trapped LC particle is less than $\pi/2$ ($\delta_{LC}=1.2<\pi/2$, which we quantify separately by analysing MM of LC in Section (\ref{exp_methods}), see Fig.~(\ref{ret})), the same helicity (input helicity) component dominates at the off-axis position as shown in Figs.~\ref{Stokes vector} (a) and (e). This means the central trap and off-axis trap particles will spin in the same direction (as predicted by Eq.~(\ref{eq3})).
    \item When the linear retardance value of the centrally trapped LC particle equals $\pi/2$ ($\delta=\pi/2$), the coefficient of the first ($cos(\delta/2)$) and second ($sin(\delta/2)$) terms of Eq.~(\ref{eq3}) equally contribute, so that the off-axially trapped LC particles will spin either clockwise or counter-clockwise, depending on their position (see Fig.~\ref{schematic_machine} (b)). Note that the values of $\sigma_{+}$ and $\sigma_{-}$ are equal but spatially separated, as shown in Figs.~\ref{Stokes vector} (b) and (e). Thus, if an off-axially trapped particle lies in the $\sigma_{+}$ region, it will spin clockwise, while if it is trapped in the $\sigma_{-}$ region, it will spin counter-clockwise. 
    \item Finally, when the linear retardance ($\delta_{LC}$) value of the centrally trapped LC particle exceeds $\pi/2$ ($\delta=2>\pi/2$), the coefficient of the second term (representing the opposite helicity) of Eq.~(\ref{eq3}) ($sin(\delta/2)$) dominates, so that the LC particle orbiting due to the fluid flow induced by the centrally trapped particle  will spin in the opposite direction from the latter (see schematic Fig.~\ref{schematic_machine} (a)). We also demonstrate this numerically in Figs.~\ref{Stokes vector} (c) and (f), where it is evident that the opposite helicity component dominates at off-axis positions.
\end{enumerate} 

Thus, our numerical simulation results align with theoretical predictions, demonstrating the composite effect of tight focusing and anisotropic medium according to Eqs.~(\ref{eq2}) and (\ref{eq3}). In what follows, we describe our experiments and observations.

\begin{center}
\begin{figure}[!t]
\includegraphics[width=0.5\textwidth]{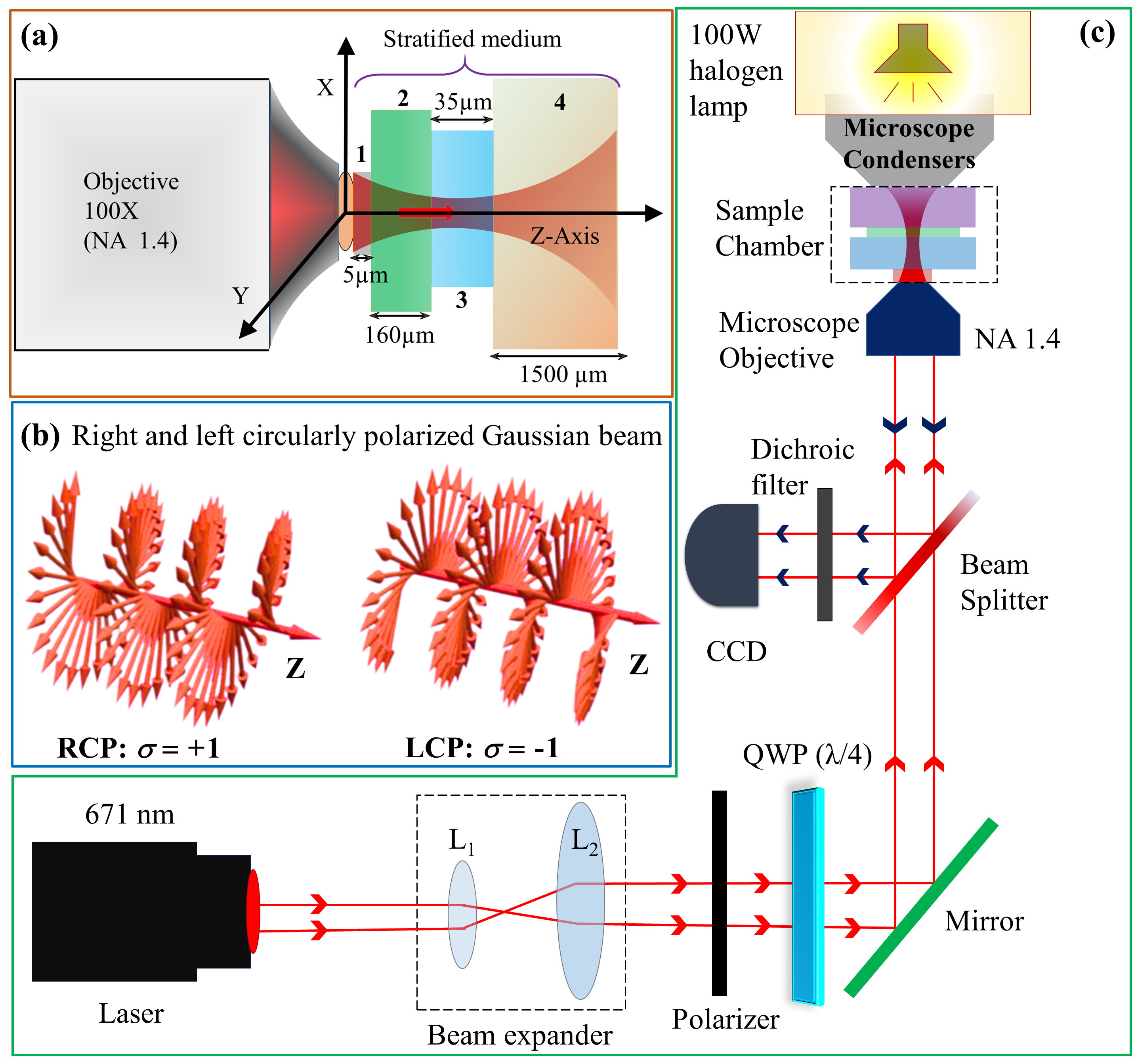}
\caption{Schematic diagram of our experimental setup. (a) Illustration of the stratified medium used in both the numerical simulation and experiment. (b) Representation of the right and left circulation of the electric field, corresponding to RCP and LCP Gaussian beams. (c) Ray diagram showing the tight focusing of circularly polarized Gaussian beams in optical tweezers.}
\label{schematic}
\end{figure}
\end{center}

\subsection{Experimental results}

\subsubsection{Micromotor 1: Primary and secondary micromotors produced by helicity and spin flow of light}\label{micro_1}

\begin{center}
\begin{figure}[!h]
\includegraphics[width=0.5\textwidth]{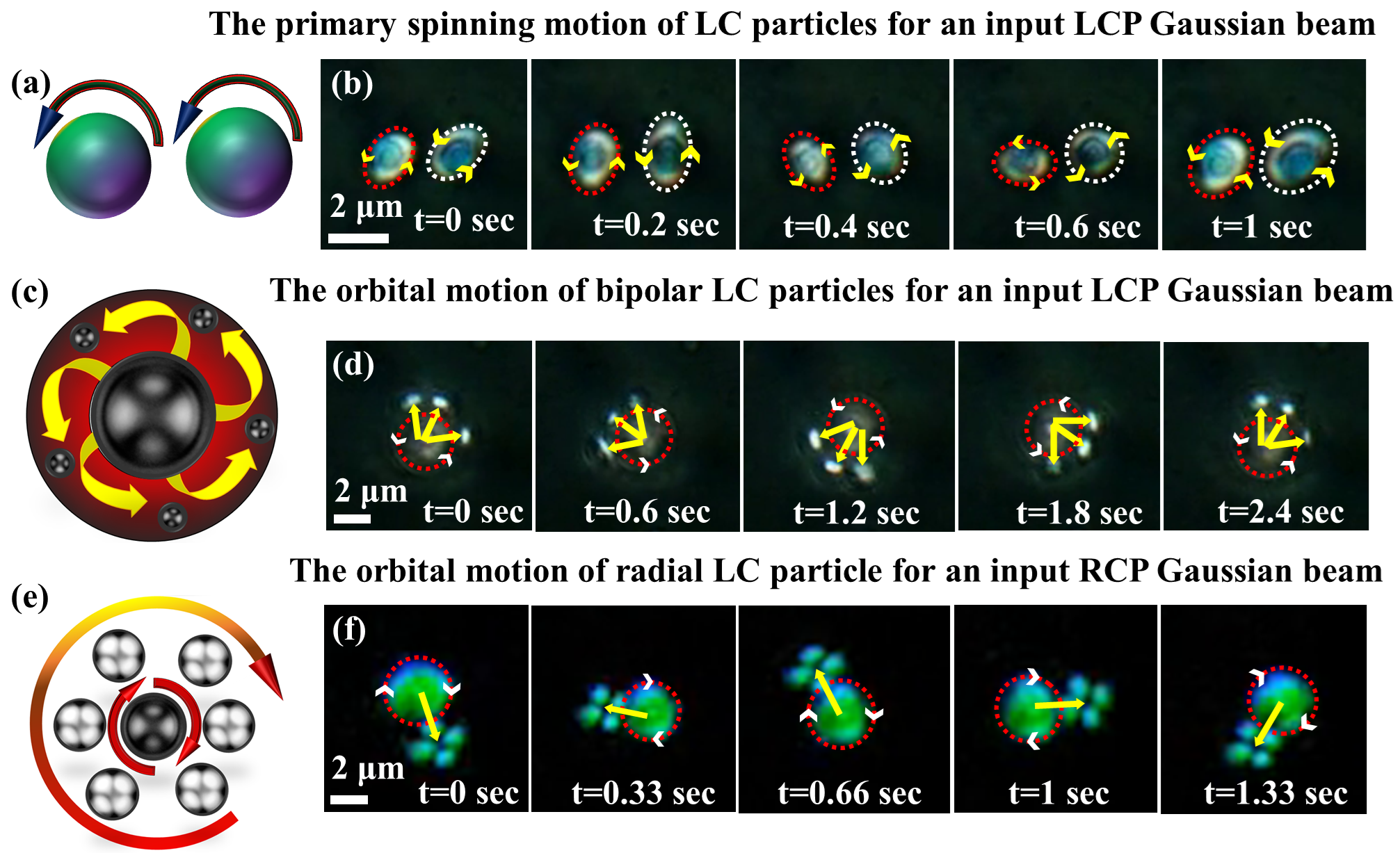}
\caption{(a) Cartoon representation of the primary micromotor. (c) and (e) Cartoon representations of the secondary micromotor. Experimental results: (b) Time-lapsed frames from Video1 showing particles within the red and white dotted ellipses spinning counterclockwise, representing the primary micromotors (due to the primary action of input helicity) under cross-polarization for input LCP light. (d) and (f) Time-lapsed frames from `Video2' and `video3', respectively, showing the orbital motion of bipolar (d) and radial (f) LC particles at off-axis positions, indicated by yellow arrow lines, representing the secondary micromotors under cross-polarization for input LCP and RCP Gaussian beams (these videos `Video1', `Video2' and `video3' were recorded using a CCD during the experiments).}
\label{micromotor1}
\end{figure}
\end{center}

The schematic in Fig.~\ref{micromotor1}(a), (c), and (e) illustrates a micromotor system driven by the helicity of light and spin flow (or fluid flow) produced by rotation of the primary micromotor. In the first case, we show the results of two spatially separated (or independent) micromotors spinning due to the LSAM present in a tightly focused circularly polarized Gaussian beam. To obtain quantitative proof of the effects of the LSAM (or helicity), we utilized a cross-polarization scheme, as detailed in the experimental section (\ref{exp_methods}). As the LC particles spin due to helicity transfer, the intensity pattern across their surface continuously changes throughout the time-lapse images. Thus, the rotation in the xy-plane (i.e., along the beam propagation z-axis) implies that the nature of the SAM is longitudinal. In Fig.~\ref{micromotor1}(b) (time-lapsed images from `Video1'), the LC particles within the red and white dotted ellipses spin counterclockwise (indicated by the small yellow arrows on the dotted ellipses) for input LCP light. Both LC particles are trapped in the spherically aberrated intensity profile of a tightly focused circularly polarized Gaussian beam under the mismatched condition of RI 1.814 as shown in Fig.~\ref{intensity}(a). Due to primary helicity ($\sigma_{-}=-1$) transfer from the input beam, the spatially separated LC particles spin in the same counterclockwise direction.

In the second case, the spinning motor at the beam center (on-axis) induces a fluid (or spin) flow in its surroundings, causing smaller motors to orbit in the direction of this flow. An LCP-input Gaussian beam generates a counterclockwise fluid flow, while an RCP-input Gaussian beam produces a clockwise flow. In Fig.~\ref{micromotor1}(d) (time-lapsed images from `Video2'), the central spinning motor within the red dotted circle spins counterclockwise, and the off-axis motors orbit due to the induced fluid flow as indicated by the yellow arrow lines. Both the spinning and orbiting LC particles are bipolar and have asymmetrically (bipolar) oriented anisotropic directors. Therefore the orbiting particles also spin due to secondary helicity transfer through the centrally trapped (on-axis) LC particle. Further details will be discussed in subsequent section (\ref{micro_2}). However, in Fig.~\ref{micromotor1}(f), the orbiting micromotors do not spin (as shown in the time-lapse images from `Video3'). This occurs because the primary micromotor is bipolar, while the off-axis secondary micromotors exhibit a radial configuration of their directors, resulting in the orbiting micromotors not spinning due to the spherically symmetric distribution of their anisotropy axis.

\subsubsection{Micromotor 2: Simultaneous revolution and rotation of secondary micromotors due to q-plate action of the primary micromotor}\label{micro_2}

\begin{center}
\begin{figure}[!h]
\includegraphics[width=0.5\textwidth]{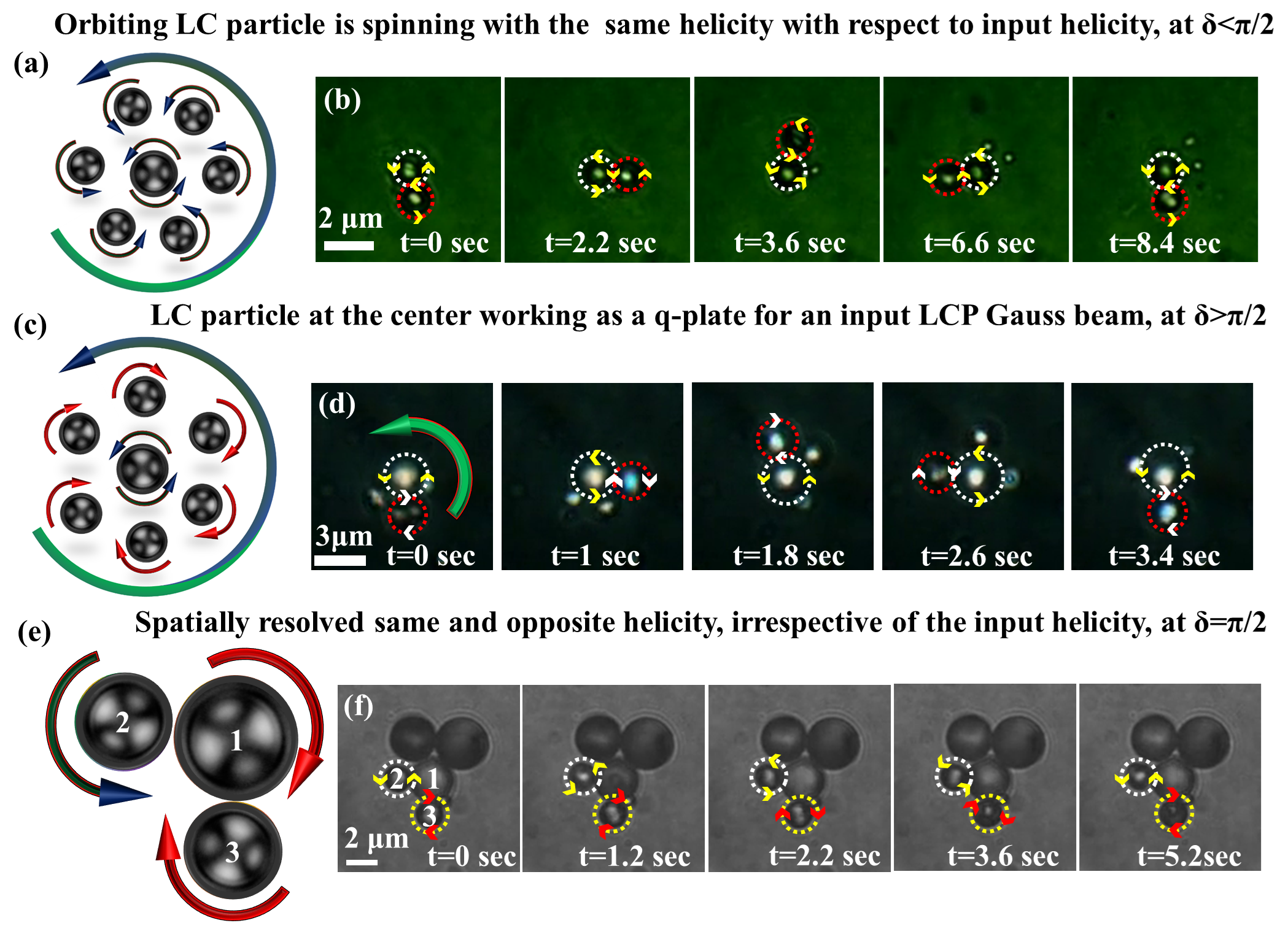}
\caption{(a) and (c) Cartoon representations of the spinning motion of the primary micromotor at the beam center (on-axis) and the spinning and orbiting motion of the secondary micromotor at the off-axis position. (e) Cartoon illustration of the spatially resolved opposite spinning motion of the secondary micromotors at off-axis positions. Experimental results: (b) and (d) Time-lapsed frames from `Video4' and `Video5' showing simultaneous primary and secondary micromotor motion for input LCP light. In both cases, the on-axis trapped particle (white dotted circle) spins counterclockwise due to primary helicity transfer, while the off-axis trapped particle (red dotted circle) orbits due to fluid flow and spins with (b) the same helicity and (d) opposite helicity due to secondary helicity transfer. (f) Time-lapsed frames from `Video6' showing spatially resolved secondary micromotor motion. The LC particles in the white and yellow dotted circles (off-axis) spin counterclockwise and clockwise, respectively, due to spatially resolved secondary helicity transfer for input RCP light. These videos- `Video4', `Video5', and `Video6' were recorded using a CCD during the experiments.}
\label{micromotor2-3}
\end{figure}
\end{center}

We now study instances where the particles (or secondary micromotor) orbiting around the primary micromotor also spin. Indeed, such simultaneous spinning and orbiting (planetary-like) motion of objects on a microscopic scale is rarely observed. In our case, the combined effects of tight focusing and the anisotropic medium of the LC particle lead to the generation of both same and opposite helicity (or secondary helicity) in the orbiting particles, as we demonstrated in our numerical simulations shown in Figs.~\ref{Stokes vector}(a)-(f). The field emerging from the central LC particle couples with the outer (off-axis) LC particles at different \(z\)-planes due to the diverging nature of the field, causing them to spin in the same and opposite direction as a consequence of SOI. Consequently, we observe simultaneous primary and secondary micromotors (the planetary-like motion of objects on a microscopic scale) driven by SOI for input LCP or RCP light. Cartoon representations of the spinning motion of the primary micromotor at the beam center (on-axis) and the orbiting and spinning motion of the secondary micromotor off-axially (planetary-like motion) are shown in Figs.~\ref{micromotor2-3} (a) and (c) for an input LCP  Gaussian beam. In both figures, the outer deep blue circular arrows illustrate the direction of the fluid flow (or spin flow) of the centrally spinning primary micromotor, while the circular arrows associated with the secondary micromotors indicate their spinning directions. We now demonstrate this experimentally, in Fig.~\ref{micromotor2-3}(b) (time-lapsed images taken from `Video4'), the primary micromotor (on-axis), depicted within the white dotted circle, spins counter-clockwise (indicated by the small yellow arrows) at very high speed for an input LCP Gaussian beam. This can be confirmed by observing the direction of the surrounding orbiting secondary micromotors (or by reducing the power of the input beam). Additionally, the outer (off-axis) particle, depicted within the red dotted circle, orbits due to the fluid flow and spins in the same counter-clockwise direction (indicated by the small yellow arrows) relative to the central spinning particle, which is, however, trapped at a different z-plane. This occurs because the value of linear retardance ($\delta_{LC}$) of the primary micromotor is less than $\pi/2$ (i.e., $\delta_{LC}=1.2<\pi/2$). Consequently, the coefficient of the first term (same helicity) of Eq.~(\ref{eq3}) ($\cos(\delta/2)$) dominates, causing the orbiting secondary micromotor to spin in the same direction as the central spinning particle, as shown in Figs.~\ref{Stokes vector} (a) and (d).

In another case, though, when the linear retardance ($\delta_{LC}$) value of the primary micromotor exceeds $\pi/2$ ($\delta=2>\pi/2$), the coefficient of the second term (opposite helicity) of Eq.~(\ref{eq3}) ($\sin(\delta/2)$) dominates, causing the orbiting secondary micromotor to spin in the opposite direction with respect to the primary, as suggested by our theory and numerical simulations (shown in Figs.~\ref{Stokes vector} (c) and (f)).  In Figs.~\ref{micromotor2-3} (d) (time-lapsed images taken from `Video5'), the primary micromotor depicted in the white dotted circle spins counterclockwise (indicated by the green arrow) at very high speeds, which can again be confirmed by observing the direction of the surrounding orbiting particles (or by reducing the power of the input beam) for an input LCP Gaussian beam. However, the secondary micromotors depicted in the red dotted circle orbit in the same direction due to the fluid (or spin) flow and spin in the opposite direction relative to the primary micromotor, again at different z-planes. In fact, in this case, the primary micromotor acts as a q-plate due to the azimuthally varying orientation of the anisotropic LC molecules directors \cite{de1993physics,marrucci2006optical,brasselet2009optical}. When the RCP/LCP Gaussian beam passes through the q-plate, it converts to an LCP/RCP optical vortex beam due to the spin-orbit conversion \cite{marrucci2006optical,kumar2024rectangular}. In all these cases, we determine the direction of motion using our cross-polarization detection mechanism (details provided in Section (\ref{exp_methods})), which reveals the spinning direction of the micromotors.

\subsubsection{Micromotor 3: Spatially controlled rotation of secondary micromotors driven by secondary helicity through primary micromotor}\label{micro_3}

For another class of secondary micromotors, we observe the generation of spatially resolved same and opposite helicity in their motion, depending on the off-axis spatial position where they are trapped around the primary motor. This is due to the fact that these micromotors have a linear retardance around $ \delta_{LC} \sim \pi/2$. In Fig.~\ref{micromotor2-3}(e), a cartoon representation illustrates the spatially resolved clockwise (depicted particle No. 3) and counterclockwise (depicted particle No. 2) motions of the secondary micromotors trapped off-axis with respect to the central trapped particle (depicted particle No. 1). The circular arrows indicate the directions of the spinning motions of the individual secondary micromotors. In Fig.~\ref{micromotor2-3}(f) (time-lapsed images taken from `Video6'), the secondary micromotors trapped off-axis, depicted in the white and yellow dotted circles, spin in the counterclockwise (opposite helicity) and clockwise (same helicity) directions, respectively. This simultaneous clockwise and counterclockwise motion of the secondary micromotors is possible due to the isotropic inhomogeneous distribution of both $\sigma=+1$ and $\sigma=-1$ secondary helicities of light off-axially, as shown in Figs.~\ref{Stokes vector} (b) and (e). The centrally trapped large primary micromotor (depicted particle No. 1) also spins slowly in a clockwise direction, which we verified by observing through the microscope eyepiece (it followed the input helicity of light). The other two large secondary micromotors are trapped off-axis due to the intensity gradient. One of them also spins very slowly, but its direction of spinning motion cannot be determined by viewing `Video6'. Consequently, we probe the direction using our cross-polarization detection mechanism, which reveals the direction of the spinning motion of the primary or secondary micromotors. Thus, both the same and opposite spins are obtained for different micromotors at different spatial locations near the central micromotor.

\section{Experimental methods and sample preparation}\label{exp_methods}

\subsection{Experimental setup}

The schematic and details of our optical tweezers setup are provided in Fig.~\ref{schematic} (a)-(c). We use a conventional optical tweezers configuration consisting of an inverted microscope (Carl Zeiss Axiovert.A1) with an oil-immersion 100X objective (Zeiss, NA 1.4) and a solid-state laser (Lasever, 671 nm, 350 mW) coupled to the back port of the microscope. A wire-grid polarizer generates linearly x-polarized light from the partially polarized Gaussian beam of the laser. The fast axis of a quarter-wave plate (QWP), centered at 671 nm, is oriented at \(45^{\circ}\) and \(135^{\circ}\) to the input beam axis to convert linearly x-polarized into right and left circularly polarized light, respectively. For the probe particles, we use nematic liquid crystals (LC), which are optically anisotropic and birefringent, allowing the transfer of angular momentum (spin) from the beam to the particles \cite{beth1936mechanical,sandomirski2004highly,kumar2024probing}. This setup facilitates probing the effects of longitudinal SAM density. We then couple the circularly polarized (RCP/LCP) Gaussian beam into the microscope, tightly focusing it into the stratified medium described earlier. Throughout the experiment, we use mismatched coverslips (RI 1.814) to enhance the effects of spherical aberration near the focal region. The influence of the stratified medium increases the spatial extent of the beam profile \cite{roy2013controlled,kumar2024probing}, which is crucial for trapping large or multiple small birefringent LC particles simultaneously to probe the effect of LSAM. The sample chamber, formed by a cover slip and glass slide sandwiched together, contains approximately 20-30 $\mu$L of the aqueous dispersion of LC micromotors/particles. The mean size of the LC droplets is 2-8 $\mu$m with a standard deviation of 20\%. We collect both the forward-transmitted light from the microscope lamp and the back-reflected light from the LC particles to characterize their spin or orbital motion. The LSAM transfer to particles trapped at the beam center and off-axis position (annular intensity ring; see Fig.~\ref{intensity} (a)) is optimized by varying the $z$-focus of the microscope objective. Note that we use a cross-polarization scheme to discern the effect of LSAM, referred to as ``yaw rotation," as previously employed in Ref.~\cite{vaippully2021continuous}, by placing crossed linear polarizers at the input and output of the microscope. Due to cross-polarization, polarization-dependent intensity lobes appear across the surface of the LC particle according to its scattering properties. 


\subsection{Sample Preparation}

For the probe particles, we use LC Colloids, which are optically anisotropic and birefringent, allowing the transfer of SAM from the beam to the particles \cite{beth1936mechanical,de1993physics,sandomirski2004highly,kumar2024probing}. These colloids are derived from Nematic Liquid-Crystal 5CB (4'-Pentyl-4-biphenylcarbonitrile). The dispersed solution of LC Colloids is prepared by mixing $5\mu l$ of 5CB and $100ml$ of deionized water. The sample is then centrifuged for 2 to 5 minutes. This process yields a concentrated dispersed solution of LC Colloids (or particles/droplets) with radii ranging from $1\mu m$ to $5\mu m$. The concentration of LC Colloids can be adjusted as needed by diluting with deionized water. The resulting dispersed solution exhibits various configurations of the LC director, including radial, pre-radial, twisted radial, monopolar, bipolar, twisted bipolar, sunset and many more structures\cite{prishchepa2008optical,de1993physics,muvsevivc2017liquid}. However, by adding $100-150\mu l$ of polyvinyl alcohol (PVA) to the solution, we can specifically obtain bipolar configurations of LC particles, which we verified using the cross-polarization technique for imaging \cite{de1993physics,brasselet2009optical,kumar2024spatially}.

The dielectric constants ${\epsilon}_{||}$ and ${\epsilon}_{\perp}$ are related to the ordinary and extraordinary refractive indices ${n}_{||}$ and ${n}_{\perp}$ of the birefringent particle through the relationships ${n}^2_{||}= {\epsilon}_{||}$ and ${n}^2_{\perp}= {\epsilon}_{\perp}$. \cite{sandomirski2004highly}. The optical birefringence $\Delta n$ value ranges from 0.1 to 0.4 for the nematic LC particles, resulting in a linear retardance ($\delta_{LC}$) value ranging from 0.5 to 3 \cite{muvsevivc2017liquid,sandomirski2004highly}.

\subsection{Mueller matrix measurement procedure}

To determine the Mueller matrix of LC particles, we employ a polarization state generator (PSG) consisting of a polarizer and a quarter-wave plate at the input of the microscope, and a polarization state analyzer (PSA) consisting of a quarter-wave plate and a polarizer at the output of the microscope. The PSG generates six different linear and circular polarization states. For this purpose, the collimated white light from the microscope’s inbuilt illumination source (a 100W halogen lamp) passes through the PSG unit, which includes a rotatable linear polarizer and a quarter-wave plate corresponding to 671- $nm$. The light is then directed onto our sample of birefringent LC, and the scattered light from the sample is collected by the microscope objective (Carl Zeiss Axiovert.A1, NA 1.4). Subsequently, the polarization of the scattered light is analyzed by the PSA unit, which comprises a quarter-wave plate and a linear polarizer. For each polarization state generated by the PSG, six measurements are taken with the PSA, yielding a total of 36 polarization-resolved measurements. By analyzing these measurements according to Table (\ref{table:Mueller}), we can construct the 4x4 Mueller matrix \cite{gupta2015wave, nayak2023spin}.

\begin{center}
\begin{figure}[!t]
\includegraphics[width=0.5\textwidth]{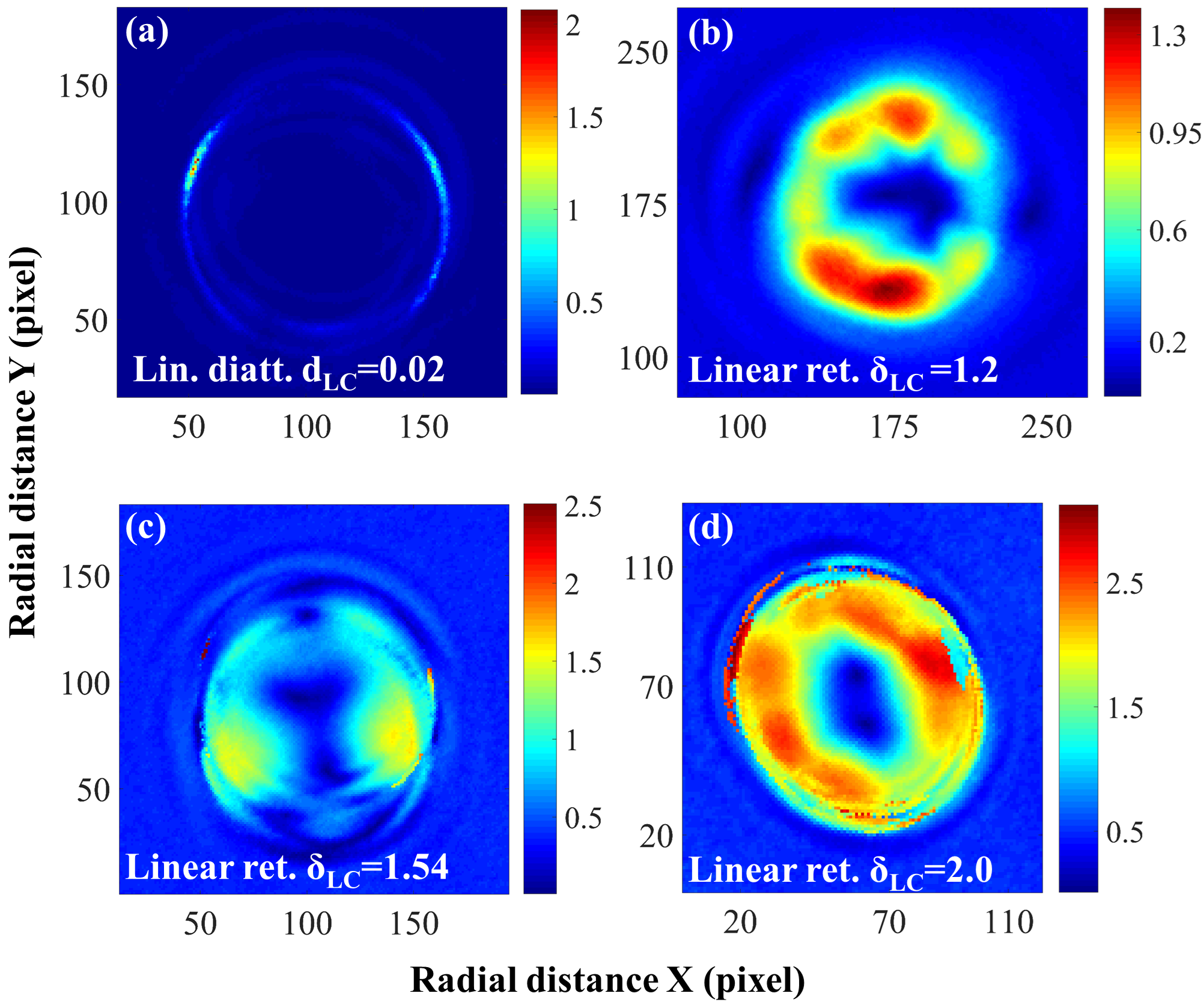}
\caption{ The polarization parameters of LC particles (a)linear diattenuation $d_{LC} \approx 0.02$. The linear retardance (b) $\delta_{LC} \approx 1.2 < \pi/2 $ (c) $\delta_{LC} \approx 1.57 \lessapprox \pi/2 $ (d) (c) $\delta_{LC} \approx 2.0 > \pi/2 $.}
\label{ret}
\end{figure}
\end{center}

\begin{table}
\centering
\scalebox{0.73}{$
\begin{tabular}{|c|c|c|c|} 
\hline
HH+HV+VH+VV & HH+HV$-$VH$-$VV & PH+PV$-$MH$-$MV & RH+RV$-$LH$-$LV\\
\hline
HH$-$HV+VH$-$VV & HH$-$HV$-$VH+VV & PH$-$PV$-$MH+MV & RH$-$RV$-$LH+LV\\
\hline
HP+VP$-$HM$-$VM & HP$-$VP$-$HM+VM & PP$-$PM$-$MP+MM & RP$-$RM$-$LP+LM\\
\hline
HR+VR$-$HL$-$VL & HR$-$VR$-$HL+VL & PR$-$PL$-$MR+ML & RR$-$RL$-$LR+LL \\
\hline
\end{tabular}$}
\caption{Scheme for construction of 4X4 Mueller matrix using 36 polarization-resolved projective measurements. Here, the first letter represents the input polarization state, and the second letter stands for the analyzer or the projected polarization state. The states are defined as $I_H(horizontal)$, $I_V(vertical)$, $I_P (+45\deg)$, $I_M(-45\deg)$, $I_L$ left circular polarized $(LCP)$, $I_R(RCP)$  }
\label{table:Mueller}
\end{table}

Using the Lu-Chipman (or polar) decomposition method \cite{xing1992deterministic, lu1995interpretation, nayak2023spin}, we quantified the polarization parameters, specifically linear diattenuation (\( d_{LC} \)) and linear retardance (\( \delta_{LC} \)), of the LC particles, as illustrated in Fig.~(\ref{ret}). We collected multiple sets of  experimental data and observed that our sample exhibited a relatively constant diattenuation value of approximately \( d \approx 0.02 \), as shown in Fig.~\ref{ret}(a). However, the linear retardance (\(\delta_{LC}\)) varied between 1 and 3, which we categorized into three distinct groups: \(\delta_{LC} < \pi/2\), \(\delta_{LC} \approx \pi/2\), and \(\delta_{LC} > \pi/2\). Figures \ref{ret}(b), (c), and (d) display the non-uniform distribution of linear retardance values for these three groups, respectively. Consequently, the LC particles are characterized by low diattenuation, indicating they are non-absorbing and function as retarders, introducing a phase difference between two orthogonal linear polarization states.

\section{Conclusion}

In conclusion, we have introduced several types of optical micromotors driven by the SOI of light, capable of even mimicking planetary-like motion. Our experiments with birefringent LC micromotors demonstrate that these micromotors can efficiently convert the angular momentum of light into controlled rotational motion without the need for traditional mechanical components. Thus, we have shown, numerically and experimentally, that the composite effect of tight focusing and an anisotropic medium leads to the breaking of helicity (RCP/LCP into LCP/RCP) as a fundamental consequence of the SOI of light in a non-paraxial regime. This results in the manifestation of both helicity components in the rotational motion of particles, which we describe as opposite helicity conversion or spin-to-spin conversion of light mediated by the SOI effect. We have demonstrated three different micromotors: a) Micromotor pair rotating in the direction of helicity of the input light, b) secondary micromotors orbiting around a primary motor due to spin (or fluid) flows induced by the latter. The orbiting micromotors may or may not spin, depending on their LC properties, with the latter mimicking a star-planet system, c) spatially resolved spin of secondary micromotors. Importantly, the direction of spin of the secondary micromotors entirely depends on the linear retardance of the primary motors. This work not only advances the understanding of light-matter interactions at the microscale but also paves the way for the development of next-generation micromotors and optical manipulation techniques. Future studies could optimise the design and control of these micromotors, exploring their potential in targeted applications such as drug delivery, optical microswitches, microfabrication, and the generation of complex fluid dynamics.

The authors acknowledge the SERB, Department of Science and Technology, Government of India (Project No. EMR/2017/001456). R.N.K. acknowledges IISER Kolkata for providing an IPh.D research fellowship.

\appendix

\section{Theoretical Calculations}\label{App_A}

We use the Debye-Wolf theory to analyze the SOI of light in the non-paraxial regime. In this approach, the incoming collimated Gaussian beam is decomposed into a superposition of plane waves, each associated with a distinct spatial harmonic component (k-vector). The Debye-Wolf integral for a tightly focused fundamental Gaussian beam passing through a high NA objective lens can be expressed as \cite{Novotny2012,roy2013controlled,richards1959electromagnetic}

\begin{eqnarray}
E(\rho,\psi,z)&=&  i\frac{kfe^{-ikf}}{2\pi}\int_0^{\theta_{max}}\int_0^{2\pi}
E_{res}(\theta,\phi)e^{ikz\cos\theta}\nonumber\\& \times & e^{ik\rho\sin\theta\cos(\phi-\psi)}
sin(\theta)\>d\theta d\phi,
\label{Eqa_1}
\end{eqnarray}

\begin{figure}[!t]
    \centering
    \includegraphics[width=0.5\textwidth]{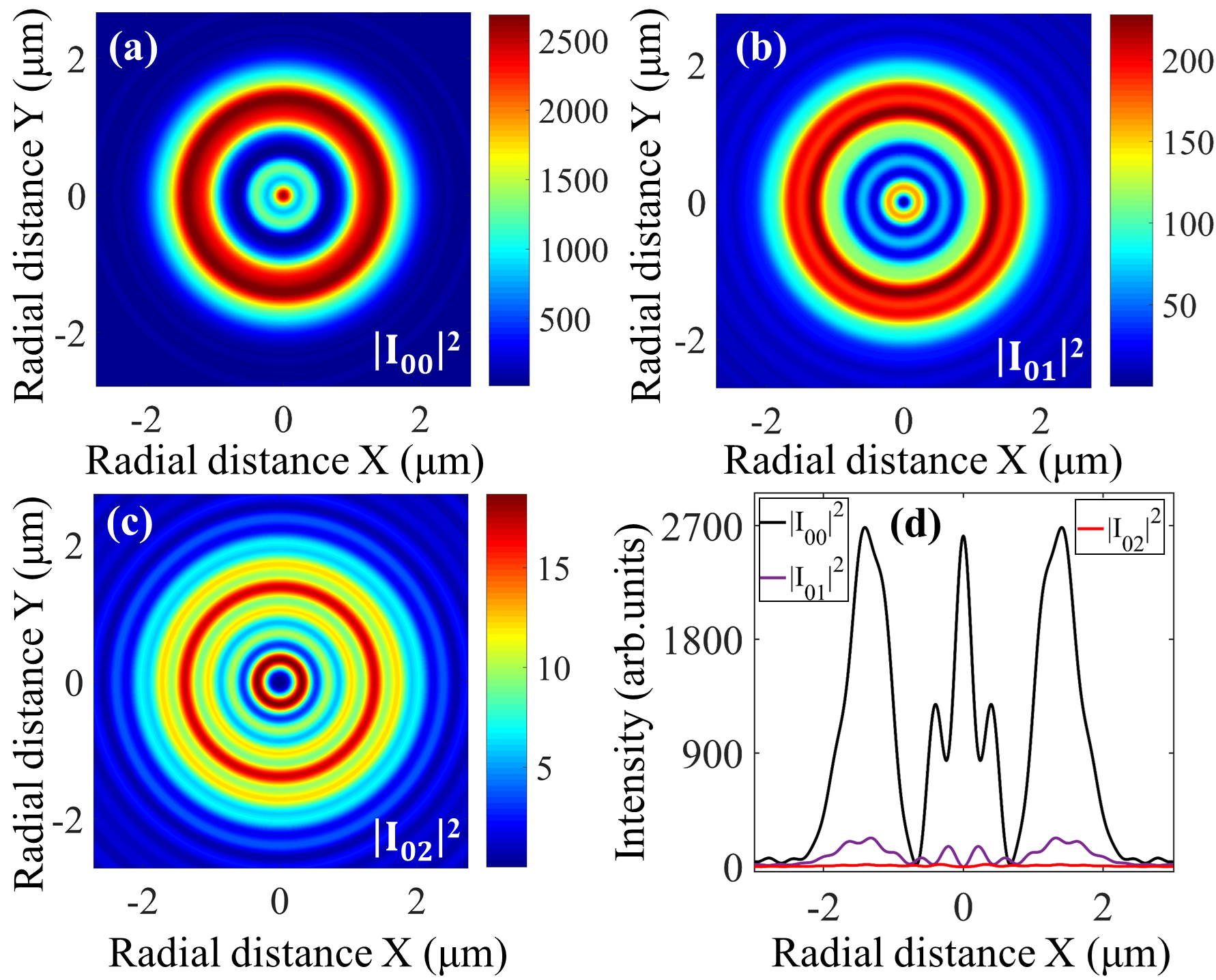}
    \caption{(a), (b), and (c) Numerical simulation of the intensity corresponding to Debye–Wolf (or diffraction) integrals \( I_{00} \), \( I_{01} \), and \( I_{02} \) at \( z = 2 \, \mu \text{m} \) away from the focus for a refractive index (RI) of 1.814, respectively. (d) Comparison of the strength of the Debye–Wolf integrals at the on-axis (beam centre) and off-axis positions in the intensity distribution.}
\label{Debye–Wolf}
\end{figure}
 
Here, \( E_{\text{res}}(\theta, \phi) \) represents the resultant electric field in the RI stratified medium, comprising a mixture of both forward- and backward-propagating waves, \( f \) is the focal length of the lens, \( k \) is the wave vector, and \( \theta_{\max} = \sin^{-1}(\text{NA} / n) \) is the maximum angle determined by the NA of the objective lens. \( n \) denotes the refractive index of the medium, while \( \theta \) and \( \phi \) refer to the polar angle relative to the z-axis and the azimuthal angle relative to the x-axis, respectively, in the cylindrical (or spherical) coordinate system.

The angular spectrum method operates in the frequency domain, where it calculates the Fourier transform (FT) of the input field \( E_{\text{inc}} \) and then multiplies it by a transfer function. The desired output field \( E_{\text{res}} \) is then obtained by taking the inverse FT. The transfer function accounts for the transformation from cylindrical to spherical coordinates. Before focusing, the incoming collimated Gaussian beam exhibits cylindrical symmetry. After tight focusing through an aplanatic lens or a high NA objective lens, the beam follows spherical symmetry. Therefore, at the transition from the paraxial to the non-paraxial regime, a transfer function representing the coordinate transformation is required.

The transfer function is given by \( A = R_z(\phi) T R_y(\theta) R_z(-\phi) \), where \( R_z \) and \( R_y \) are \( \text{SO}(3) \) rotation matrices. Since the stratification of the medium causes the field propagation to depend on the input polarization, the Fresnel transmission coefficients \( T_s \) and \( T_p \) and the Fresnel reflection coefficients \( R_s \) and \( R_p \) are incorporated, considering both \( s \)- and \( p \)-polarizations ($E_{inc} = E^{s}_{inc} + E^{p}_{inc}$). For backward-propagating waves, the transfer function \( A \) is modified by replacing \( \theta \) with \( \pi - \theta \), and the Fresnel reflection coefficients \( R_s \) and \( R_p \) are used instead of the transmission coefficients. The resultant and incident electric fields are thus related through the transfer function \( A \) as

\begin{equation}
    E_{r e s}(\theta, \phi)=A E_{i n c}(\theta, \phi)
    \label{Eqa_2}
\end{equation}
where, the $T$ and $R$ matrices are given by:

\begin{equation*}
T=\left(\begin{array}{ccc}
T_{p} & 0 & 0 \\
0 & T_{s} & 0 \\
0 & 0 & T_{p}
\end{array}\right) ; R=\left(\begin{array}{ccc}
-R_{p} & 0 & 0 \\
0 & R_{s} & 0 \\
0 & 0 & -R_{p}
\end{array}\right),
\end{equation*}

Note that we have $ T_{i}^{(1, j)}=\dfrac{E_{i+}^{j}}{E_{i+}^{1}} ; R_{i}^{(1, j)}=\dfrac{E_{i-}^{j}}{E_{i+}^{1}};$. Here, $i$ specifies the polarization $(s$ and $p),~ +/-$ signifies a wave propagating forward and back-ward, respectively, and $j$ in the superscript specifies the layer of the stratified medium in which the optical tweezers (trapping laser) focus lies. We take into account the Fresnel transmission coefficients $T_{s}$ and $T_{p}$, as well as the reflection coefficients $R_{s}$ and $R_{p}$ at the interface of the stratified medium. 
For an input Gaussian TM$_{00}$ mode \cite{Novotny2012,roy2013controlled}, we have

\begin{equation}
E_{\mathrm{inc}}=E_0 \mathrm{e}^{-f^2 \sin ^2 \theta / w_0^2}
\label{Eqa_3}
\end{equation}

\begin{figure*}[!t]
    \centering
    \includegraphics[width=\textwidth]{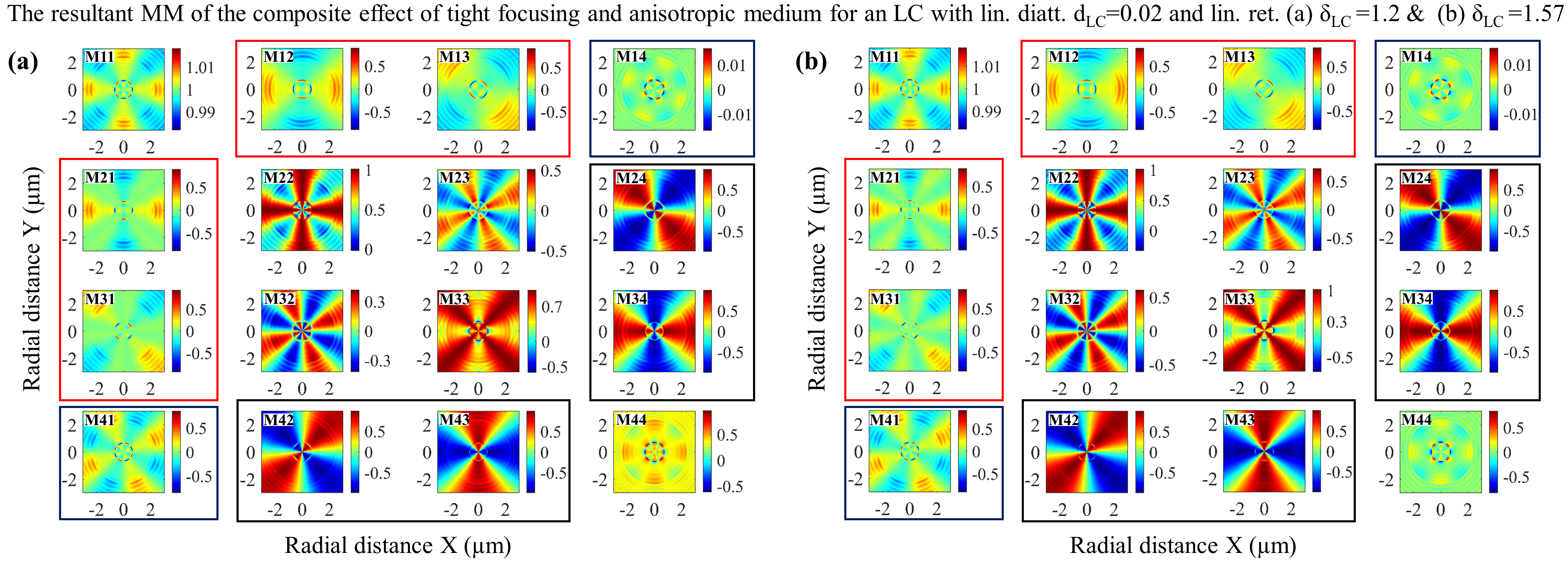}
    \caption{Numerically computed 4x4 resultant Mueller matrix for the tight focusing of a Gaussian beam on an LC particle with linear diattenuation $d_{LC}=0.02$ and retardance (a) $\delta_{LC}=1.2$ and (b) $\delta_{LC}=1.57$. The signature of the linear diattenuator is indicated by elements M${12}$, M${13}$, M${21}$, and M${31}$ (marked by red boxes). The linear retarder is indicated by elements M${24}$, M${34}$, M${42}$, and M${43}$ (marked by black boxes). The elements M${14}$ and M${41}$, which describe the Spin Hall Effect (SHE), i.e., the difference in RCP and LCP components of the electric field in the focal plane, are highlighted in blue boxes.}
\label{MM at 1.2 and 1.57}
\end{figure*}

Using equations (\ref{Eqa_1}), (\ref{Eqa_2}), and (\ref{Eqa_3}), we have calculated the 3X3 Jones matrix of tight focusing. Now, the output and the input electric fields are related through a 3X3 Jones matrix of tight focusing as:

\onecolumngrid
\begin{equation}
\left[\begin{array}{c}
E_{x}^{o} \\
E_{y}^{o} \\
E_{z}^{o}
\end{array}\right]^{Gauss}=C\left[\begin{array}{ccc}
I_{00}+I_{02} \cos 2 \psi & I_{02} \sin 2 \psi & 2 i I_{01} \cos \psi \\
I_{02} \sin 2 \psi & I_{00}-I_{02} \cos 2 \psi & 2 i I_{01} \sin \psi \\
-2 i I_{01} \cos \psi & -2 i I_{01} \sin \psi & I_{00}+I_{02}
\end{array}\right]\times\left[\begin{array}{c}
E_{x}^{i} \\
E_{y}^{i} \\
E_{z}^{i}
\end{array}\right],
\label{Eqa_4}
\end{equation}

\twocolumngrid

Here, \(E^{o}\) and \(E^{in}\) denote the output and input Jones polarization vectors, respectively, with the suffix `Gauss' indicating the Gaussian beam, and \(\psi\) representing the azimuthal angle in the cylindrical (or spherical) coordinate system. The terms \(I_{00}\), \(I_{01}\), and \(I_{02}\) correspond to the Debye–Wolf (or diffraction) integrals for the transmitted and reflected waves, as given by \cite{Novotny2012, roy2013controlled}.

\begin{widetext}
\begin{equation}
\begin{aligned}
I_{00}^t(\rho)= & \int_0^{\theta_{\max }} E_{\mathrm{inc}}(\theta) \sqrt{\cos \theta}\left(T_s^{(1, j)}+T_p^{(1, j)} \cos \theta_j\right) J_0\left(k_1 \rho \sin \theta\right) e^{i k_j z \cos \theta_j} \sin (\theta) d \theta, \\
I_{01}^t(\rho)= & \int_0^{\theta_{\max }} E_{\mathrm{inc}}(\theta) \sqrt{\cos \theta} \left(T_p^{(1, j)} \sin \theta_j\right) J_1\left(k_1 \rho \sin \theta\right) e^{i k_j z \cos \theta_j} \sin \theta d \theta, \\
I_{02}^t(\rho)= & \int_0^{\theta_{\max }} E_{\mathrm{inc}}(\theta) \sqrt{\cos \theta}\left(T_s^{(1, j)}-T_p^{(1, j)} \cos \theta_j\right) J_2\left(k_1 \rho \sin \theta\right) e^{i k_j z \cos \theta_j} \sin \theta d \theta, \\
I_{00}^r(\rho)= & \int_0^{ \theta_{\max }} E_{\mathrm{inc}}(\theta) \sqrt{\cos \theta}\left(R_s^{(1, j)}-R_p^{(1, j)} \cos \theta_j\right) J_0\left(k_1 \rho \sin \theta\right) e^{-i k_j z \cos \theta_j} \sin \theta d \theta, \\
I_{01}^r(\rho)= & \int_0^{\theta_{\max }} E_{\mathrm{inc}}(\theta) \sqrt{\cos \theta} \left(R_p^{(1, j)} \sin \theta_j\right) J_1\left(k_1 \rho \sin \theta\right) e^{-i k_j z \cos \theta_j} \sin \theta d \theta, \\
I_{02}^r(\rho)= & \int_0^{\theta_{\max }} E_{\mathrm{inc}}(\theta) \sqrt{\cos \theta}\left(R_s^{(1, j)}+R_p^{(1, j)} \cos \theta_j\right) J_2\left(k_1 \rho \sin \theta\right) e^{-i k_j z \cos \theta_j} \sin \theta d \theta,
\end{aligned}
\label{Eqa_5}
\end{equation}
\end{widetext}

The superscripts \( t \) and \( r \) denote the transmitted and reflected components, respectively. The functions \( J_0 \), \( J_1 \), and \( J_2 \) represent the zero, first, and second-order Bessel functions of the first kind, respectively.

\begin{figure*}[!t]
    \centering
    \includegraphics[width=\textwidth]{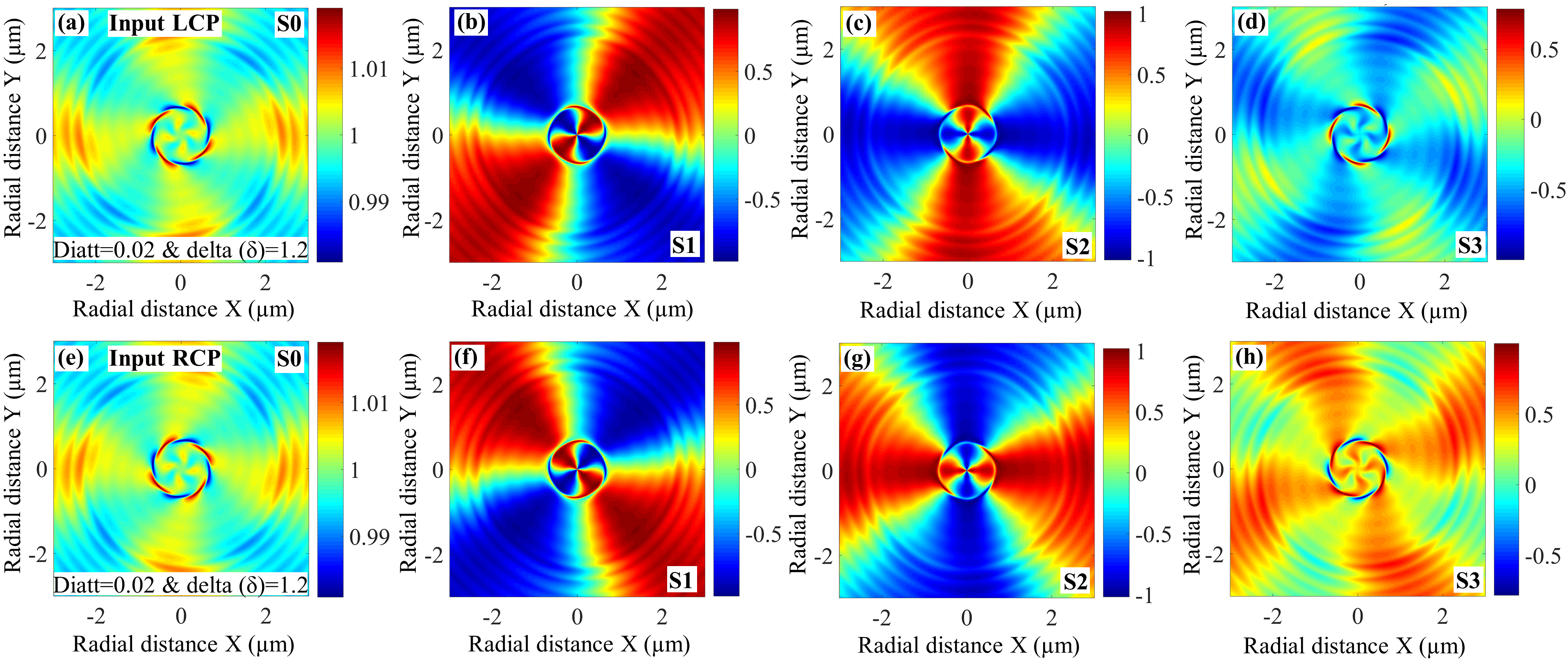}
    \caption{The resultant Stokes vector elements for tightly focused circularly polarized light on an LC particle with a fixed linear diattenuation value of 0.02 and a linear retardance of 1.2. (a)-(d) show results for left circularly polarized light, while (e)-(h) show results for right circularly polarized light. (a) and (e) depict the \( S_0 \) component, representing the normalized total intensity. (b), (c), (f), and (g) display the \( S_1 \) and \( S_2 \) components, which describe the differences in the linearly polarized components of the emerging field. (d) and (h) illustrate the \( S_3 \) component, which is proportional to the longitudinal component of spin angular momentum (LSAM) density, highlighting the spatial separation between \( \sigma = +1 \) and \( \sigma = -1 \) helicity of the focused field on LC particles.}
\label{dlta_1.2}
\end{figure*}

\begin{figure*}[!t]
    \centering
    \includegraphics[width=\textwidth]{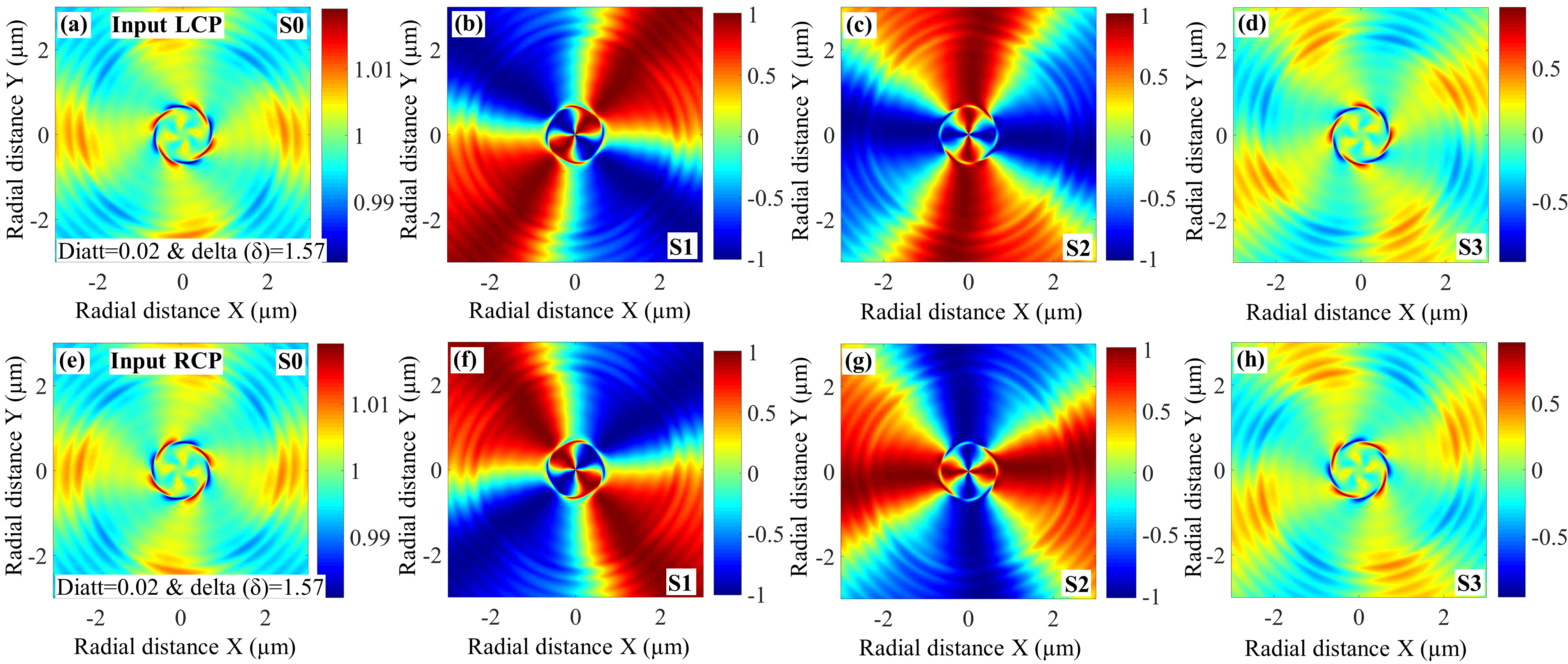}
    \caption{The resultant Stokes vector elements of tightly focused circularly polarized light on an LC particle have a fixed linear diattenuation value of 0.02 and a linear retardance of 1.57. (a)-(d) show results for left circularly polarized light, and (e)-(h) show results for right circularly polarized light. (a) and (e) The S${0}$ component represents the normalized total intensity. (b), (c), (f), and (g) The S${1}$ and S${2}$ components describe the difference in the linearly polarized components of the focused field. (d) and (h) The S${3}$ component, proportional to the longitudinal component of SAM density, describes the spatially separated $\sigma=+1$ and $\sigma=-1$ helicities of the focused field on LC particles.}
\label{dlta_1.57}
\end{figure*}

\section{Numerical Simulation}\label{APP_B}

Our simulations are performed for the tight focusing of an input circularly polarized Gaussian beam by a high NA objective lens into a stratified medium, as described in Section (\ref{Num_simulations}). In the focal plane (or near the focal plane), the electric field exhibits components not only along the transverse direction but also along the longitudinal direction, due to the transverse boundary conditions. As mentioned in Section (\ref{Num_simulations}), the electric field intensity at the center of the beam profile appears as bright spots because the zero-order Bessel function (\( J_{0} \), embedded in \( I_{00} \)), has a non-vanishing value at the origin (on-axis). However, the intensity corresponding to the longitudinal component of the electric field (\( E_z \)) is primarily concentrated at off-axis positions due to the first-order Bessel function (\( J_{1} \)), embedded in the \( I_{01} \) coefficient. In Figs.~\ref{Debye–Wolf} (a), (b), and (c), we show the distribution of the squared moduli of \( I_{00} \), \( I_{01} \), and \( I_{02} \), respectively. However, in Fig.~\ref{Debye–Wolf} (d), we compare the strength of the Debye–Wolf integrals at the on-axis (beam center) and off-axis positions. The maximum value of the squared modulus of \( I_{00} \) occurs at both the on-axis (beam center) and off-axis positions, with a value around 2700 (arb. units). In contrast, the maximum value of the squared modulus of \( I_{01} \) primarily occurs at off-axis positions, with a value around 230 (arb. units), while the squared modulus of \( I_{02} \) is negligible.

\subsection{Mueller matrix of composite effect}

The tight focusing of the beam and the bipolar variation of the anisotropy axis of the LC particles together create a composite effect, as described in Sec.~(\ref{res_dis}). Here, we present the resulting Mueller matrix (MM) for the case \(\delta < \pi/2\) and \(\delta_{LC} = \pi/2\). The results for \(\delta < \pi/2\) and \(\delta_{LC} = \pi/2\) are shown in Fig.~\ref{MM at 1.2 and 1.57}, with the case of \(\delta_{LC} > \pi/2\) already discussed in Section (\ref{res_dis}). We then calculate the resulting Stokes vector using the equation \(\mathbf{S}^0 = M_{\text{res}} \mathbf{S}^{\mathbf{in}}\).

\subsection{Resultant Stokes Vector}

Based on the experimental results, we calculated the resultant Stokes vector elements \( S0 \), \( S1 \), \( S2 \), and \( S3 \) for input left circularly polarized (LCP) light \((S_{LCP}^{in} = \left[\begin{array}{llll} 1 & 0 & 0 & -1 \end{array}\right]^{T})\) and right circularly polarized (RCP) light \((S_{RCP}^{in} = \left[\begin{array}{llll} 1 & 0 & 0 & 1 \end{array}\right]^{T})\). These calculations were performed using the resultant Mueller matrix \( M_{\text{res}} \), which characterizes the composite effect of tight focusing of a Gaussian beam on an LC particle (\(\mathbf{S}^0 = M_{\text{res}} \mathbf{S}^{\mathbf{in}}\)). We specifically considered cases where the linear diattenuation is fixed at 0.02 and the linear retardance \( \delta_{LC} \) is set to 1.2 and 1.57, as shown in Fig.~(\ref{dlta_1.2}) and Fig.~(\ref{dlta_1.57}) respectively (with prior demonstrations of \( S3 \) component for \(\delta_{LC} > \pi/2\) in section (\ref{res_dis})).



\providecommand{\noopsort}[1]{}\providecommand{\singleletter}[1]{#1}%

\end{document}